\documentclass[10pt]{iopart}
\usepackage{graphicx}
\usepackage[normalem]{ulem}
\usepackage{iopams}
\usepackage{color}

\usepackage{pdfpages}

\usepackage{ulem}

\begin{document}

\title[G Kremer \textit{et al.}]{Dispersing and semi-flat bands in the wide band gap two-dimensional semiconductor bilayer silicon oxide}

\author{Geoffroy Kremer$^{1,2,*}$, Juan Camilo Alvarez-Quiceno$^3$, Thomas Pierron$^2$, C\'{e}sar Gonz\'{a}lez$^{4,5,**}$, Muriel Sicot$^{2,***}$, Bertrand Kierren$^2$, Luc Moreau$^2$, Julien E. Rault$^6$, Patrick Le F\`evre$^6$, Fran\c{c}ois Bertran$^6$, Yannick J. Dappe$^{7,****}$, Johann Coraux$^{8,*****}$, Pascal Pochet$^{3,******}$, and Yannick Fagot-Revurat$^{2}$}
\address{$^1$ D\'{e}partement de Physique and Fribourg Center for Nanomaterials, Universit\'{e} de Fribourg, CH-1700 Fribourg, Switzerland}
\address{$^2$ Institut Jean Lamour, UMR 7198, CNRS-Universit\'{e} de Lorraine, Campus ARTEM, 2 all\'{e}e
Andr\'{e} Guinier, BP 50840, 54011 Nancy, France}
\address{$^3$ Department of Physics, IriG, Univ. Grenoble Alpes \& CEA 38054 Grenoble France}
\address{$^4$ Departamento de F\'{i}sica de Materiales, Universidad Complutense de Madrid, E-28040 Madrid, Spain}
\address{$^5$ Instituto de Magnetismo Aplicado UCM-ADIF, E-28230 Madrid, Spain}
\address{$^6$ Synchrotron SOLEIL, Saint-Aubin, BP 48, F-91192 Gif-sur-Yvette Cedex, France}
\address{$^7$ SPEC, CEA, CNRS, Universit\'{e} Paris-Saclay, CEA Saclay, 91191 Gif-sur-Yvette Cedex, France}
\address{$^8$ Univ. Grenoble Alpes, CNRS, Grenoble INP, Institut N\'{e}el, 38000 Grenoble, France}
\address{$^{*}$ ORCiD: 0000-0003-1753-3471}
\address{$^{**}$ ORCiD: 0000-0001-5118-3597}
\address{$^{***}$ ORCiD: 0000-0002-0558-2113}
\address{$^{****}$ ORCiD: 0000-0002-1358-3474}
\address{$^{*****}$ ORCiD: 0000-0003-2373-3453}
\address{$^{******}$ ORCiD: 0000-0002-1521-973X}

\begin{abstract}
Epitaxial bilayer silicon oxide is a transferable two-dimensional material predicted to be a wide band gap semiconductor, with potential applications for deep UV optoelectronics, or as a building block of van der Waals heterostructures. The prerequisite to any sort of such applications is the knowledge of the electronic band structure, which we unveil using angle-resolved photoemission spectroscopy and rationalise with the help of density functional theory calculations. We discover dispersing bands related to electronic delocalisation within the top and bottom planes of the material, with two linear crossings reminiscent of those predicted in bilayer AA-stacked graphene, and semi-flat bands stemming from the chemical bridges between the two planes. This band structure is robust against exposure to air, and can be controled by exposure to oxygen. We provide an experimental lower-estimate of the band gap size of 5 eV and predict a full gap of 7.36 eV using density functional theory calculations.
\end{abstract}

\color{black}
%
\vspace{2pc}

%
%

\ioptwocol

\section*{Introduction}
Silicon oxide can be prepared in the form of a crystalline two-dimensional (2D) allotrope with no dangling bonds, and whose structure has been resolved starting from 2010 \cite{loffler_growth_2010,Huang2012}. Observed from the top, the atomic arrangement appears as a honeycomb lattice, while a side-view reveals that it consists of two symmetric layers of corner-sharing SiO$_4$ tetrahedrons. It was therefore called a bilayer (BL).

The (0001) surface of ruthenium is a substrate of choice for the epitaxial growth of this material. With optimized growth conditions, it is possible to obtain a highly crystalline BL silicon oxide film decoupled from its surface, \textit{i.e.} without strong covalent substrate-oxide bonds \cite{klemm_preparation_2016,kuhness2018two}. Silicon oxide's unit cell is twice larger than the surface unit cell of Ru(0001), and it coexists with a more or less dense layer of O atoms directly attached to the substrate \cite{wlodarczyk_tuning_2012}. The weak interaction between BL silicon oxide and Ru(0001) recently opened the route to the transfer of the former via a mechanical exfoliation \cite{transfert}. The square-centimeter thus-transferred silicon oxide showed excellent chemical and thermal stability, suggesting its use in a variety of applications.

Bilayer silicon oxide is expected to be a wide band gap semiconductor \cite{Gao2016}. Its intrinsic electronic and excitonic properties are yet to be investigated experimentally. What the actual size of the band gap is, which is difficult to predict \textit{a priori}, or whether this band gap is direct or indirect, are key questions. The answers will determine the potential for optoelectronic applications, and allow to position this potential with respect to another wide band gap 2D material, hexagonal boron nitride \cite{Elias2019}. Beyond potential interesting optoelectronic properties of the individual layers, BL silicon oxide may also be a valuable building-block of functional artificial van der Waals heterostructures, as a dielectric layer separating smaller band gap \cite{Fang2014} or conductive \cite{Goratchev2012} 2D materials, or alternatively, stacked with a second BL silicon oxide layer, with a twist angle producing strong electron correlation effects such as those predicted in twisted BL boron nitride \cite{Xian2019}.

In the present article, we unveil the electronic band structure of epitaxial crystalline BL silicon oxide grown on Ru(0001) using angle-resolved photoemission spectroscopy (ARPES) and discuss the origin of the bands in light of density functional theory (DFT) calculations. We find dispersing bands stemming from the hybridization between electronic orbitals of Si and O atoms in the top and bottom planes of the 2D material. Some of these bands exhibit two linear crossings, that are reminiscent of those expected in a peculiarly-stacked graphene BL. We also identify semi-flat bands that originate from the Si--O--Si bridges between the top and bottom planes. This band structure is robust, and resists exposure to air. After having identified the specific electronic bands of the substrate, that do not hybridize with those of silicon oxide, we identify the valence band maximum (VBM) of the BL. The energy of this band extremum is found to vary from -4 to -5 eV below the Fermi level (E\textsubscript{F}), depending on the amount of O atoms directly bound to the Ru substrate, below the BL silicon oxide. We conclude that the electronic band gap of the material is at least 5 eV, confirming the potential of the material for deep UV optoelectronics applications and consistent with our DFT calculations which are predicting a band gap of 7.36 eV.

\begin{figure*}
\centering
\includegraphics[scale=0.63]{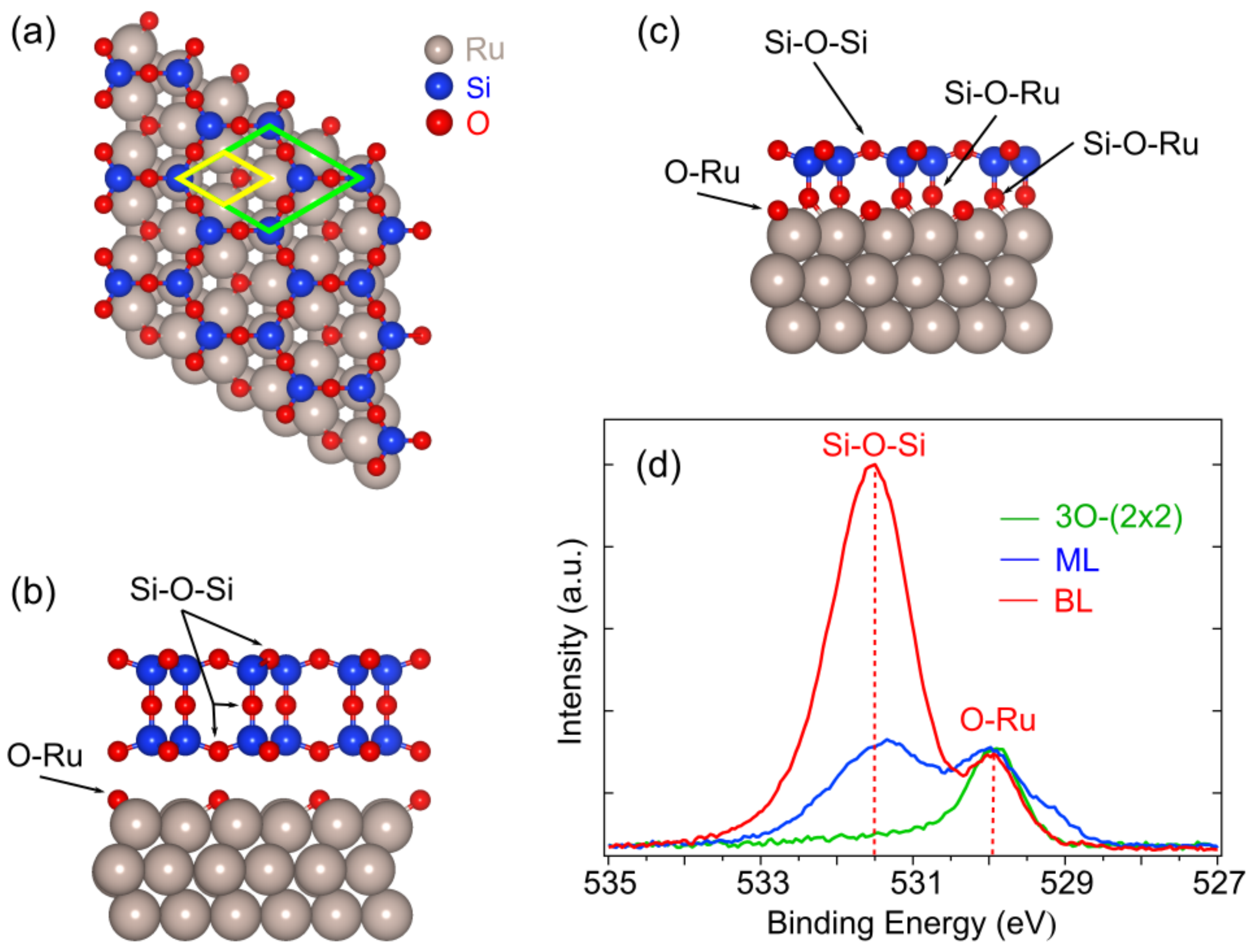}
\caption{\label{Figure1} (a,b) Schematic structure of the 2D BL silicon oxide (top and side views, respectively) on Ru(0001). 2D BL silicon oxide and Ru unit cells are indicated as green and yellow rhombuses, respectively. (c) Schematic structure (side view) of the 2D ML silicon oxide on Ru(0001). (d) High-resolution XPS spectra of the \hbox{O 1s} core level for the 3O--(2$\times$2)/Ru(0001) (green), ML silicon oxide (blue) and BL silicon oxide (red) using a photon energy of 1486.6 eV.}
\end{figure*}


\color{black}
\section*{Materials and methods}
\subsection*{Experiment}
Experiments were carried out in a ultra-high vacuum setup ($P < 1 \times 10^{-10}$\,mbar). It is equipped with low-energy electron diffraction (LEED), a monochromatized X-rays source (Al K$_{\alpha}$,  resolution better than 300 meV), monochromatized helium  source (MBS L-1), and a high energy, momentum and spin resolution photoemission analyser (DA30-L from VG--SCIENTA). ARPES measurements were carried out at room temperature with energy and angular resolutions better than 20 meV and 0.1$^{\circ}$,  respectively. The energy resolution has been determined by the fit of the Fermi edge of a polycrystalline copper sample at low temperature. Details about the experimental geometry are given in figure S1 of the supplementary information (SI). A clean Ru(0001) surface was obtained  by repeated cycles of Ar$^{+}$ sputtering and annealing up to $1400$\,K  followed by molecular oxygen exposure and flash annealing resulting in a sharp ($1\times1 $) LEED pattern (not shown here). The absence of contamination was checked by X-ray photoemission spectroscopy (XPS) and ARPES.  After cleaning, the \hbox{Ru 3d\textsubscript{5/2}} core level  exhibits a surface-related contribution at a binding energy (BE) of 279.8 eV (\hbox{figure S8(c)}). A crystalline BL silicon oxide was grown on an oxygen-covered Ru(0001) surface forming a so-called 3O--(2$\times$2) reconstruction \cite{kim_structural_1998}. The latter was obtained by exposing Ru(0001)  to $1 \times 10^{-6}$\,mbar O$_{2}$ at $625$\,K  for $10$\,min. Then silicon was evaporated, for twice as long as for the monolayer (ML) case,  using electron bombardment of a high purity Si rod (better than \hbox{99.9999 $\%$)}  under an oxygen pressure of $3 \times 10^{-7}$\,mbar at room temperature. The control of the amount of Si is the key to obtain the BL \footnote[1]{With a sufficient amount of Si indeed, Si--O--Si bridges can be formed instead of the Ru--O--Si bridges in the ML. While the latter bonds are seemingly less stable, surface energy minimisation effects might however prevail (the oxide-covered Ru(0001) surface being less energetic than the O-covered Ru(0001) surface, presumably) for smaller amount of Si deposited on the surface, promoting the formation of the ML.}, as ascertained with XPS, infrared and tunneling spectroscopies \cite{loffler_growth_2010, wlodarczyk_tuning_2012,lichtenstein_atomic_2012,yang_thin_2012,lichtenstein2012probing}. The final step was performed under $3 \times 10^{-6}$\,mbar O\textsubscript{2} at a temperature ranging from $1100$\,K to  $1300$\,K for $15$\,min followed by a slow temperature ramp, promoting crystallization (rather than the formation of the amorphous BL)  at a rate of \hbox{$10$\,K $\cdot$ min$^{-1}$} down to $300$\,K. Temperatures were measured using a pyrometer.

\subsection*{Computational details}
We used first--principles DFT \cite{dft-hk, dft-ks} calculations and the Projected Augmented Wave (PAW) method \cite{Blochl1994} as implemented in the Vienna {\it ab-initio} Simulation Package (VASP) \cite{Kresse1996}. The kinetic energy cut-off for the plane-wave expansion was 490 eV. For exchange correlation potential, we used the local spin density approximation (L(S)DA) \cite{Ceperley1980LDA} and the hybrid HSE06~\cite{Krukau2006HSE06}. We also performed calculations including van der Waals corrections through the D2 method (PBE+D2) introduced by Grimme \cite{grimme2006semiempirical}. The Ru(0001) surface was modelled by taking three atomic layers, cleaved from a face-centered cubic lattice where the crystal lattice parameter was firstly optimized. The deepest ruthenium layer was kept fixed during the relaxation of the rest of the structure. Atomic positions were optimized through the conjugate gradient algorithm until the Hellmann-Feynman forces reached the threshold of 0.1$\times$10$^{-3}$~eV/{\AA} (see supplementary .xyz file). A vacuum space of at least 12 {\AA} was employed along the $z$-direction to avoid undesired interactions between periodic layers. The Brillouin zone (BZ) was sampled with a $20\times20\times1$ k-point mesh in the self-consistent energy calculations.


\section*{Structure and growth procedure}

The BL is composed of two superimposed layers of cornersharing  SiO\textsubscript{$4$} tetrahedra forming a honeycomb-like structure (figure~\ref{Figure1}(a,b)) \cite{loffler_growth_2010,lichtenstein_atomic_2012,yang_thin_2012}. The lattice constant is 5.42~\AA, which is twice that of the Ru substrate, leading to a (2$\times$2) commensurate unit cell (figure \ref{Figure1}(a)). Contrary to the ML silicon oxide, which is connected to Ru by covalent Si--O--Ru bonds (figure~\ref{Figure1}(c)), the interaction between the BL and its substrate is dominated by weak van der Waals forces. One evidence is the absence of Si--O--Ru perpendicular vibration modes \cite{loffler_growth_2010,lichtenstein_atomic_2012,fischer_ultrathin_2015}. Another experimental proof is the absence of Si--O--Ru contributions in XPS \cite{loffler_growth_2010,wlodarczyk_tuning_2012}.

Measuring the electronic band structure with ARPES usually requires crystalline samples with well-defined crystallographic orientation. The key parameters to achieve this with BL silicon oxide are the temperature and the partial pressure during the annealing step. Heating up to 1300 K under $3 \times 10^{-6}$\,mbar O\textsubscript{2} produces a  sharp ($2 \times 2$) LEED pattern,  indicative of a good structural quality whereas a lower annealing temperature produces a diffuse ring representative of amorphous domains (\hbox{figure S3}).

The BL can be grown either on an oxygen-free underlying substrate or on an oxygen-reconstructed Ru surface (from a so-called O--(2$\times$2) to a so-called \hbox{O--(1$\times1$)}). These interfacial O atoms seem to play an important role concerning the crystallinity of the BL. Crystallinity seems optimum when the initial coverage of Ru(0001) with O is high \cite{klemm_preparation_2016}, which is what we targeted by using a large O$_2$ dose to produce a \hbox{3O--(2$\times$2)} superstructure.

\begin{figure*}
\begin{center}
\includegraphics[scale=0.63]{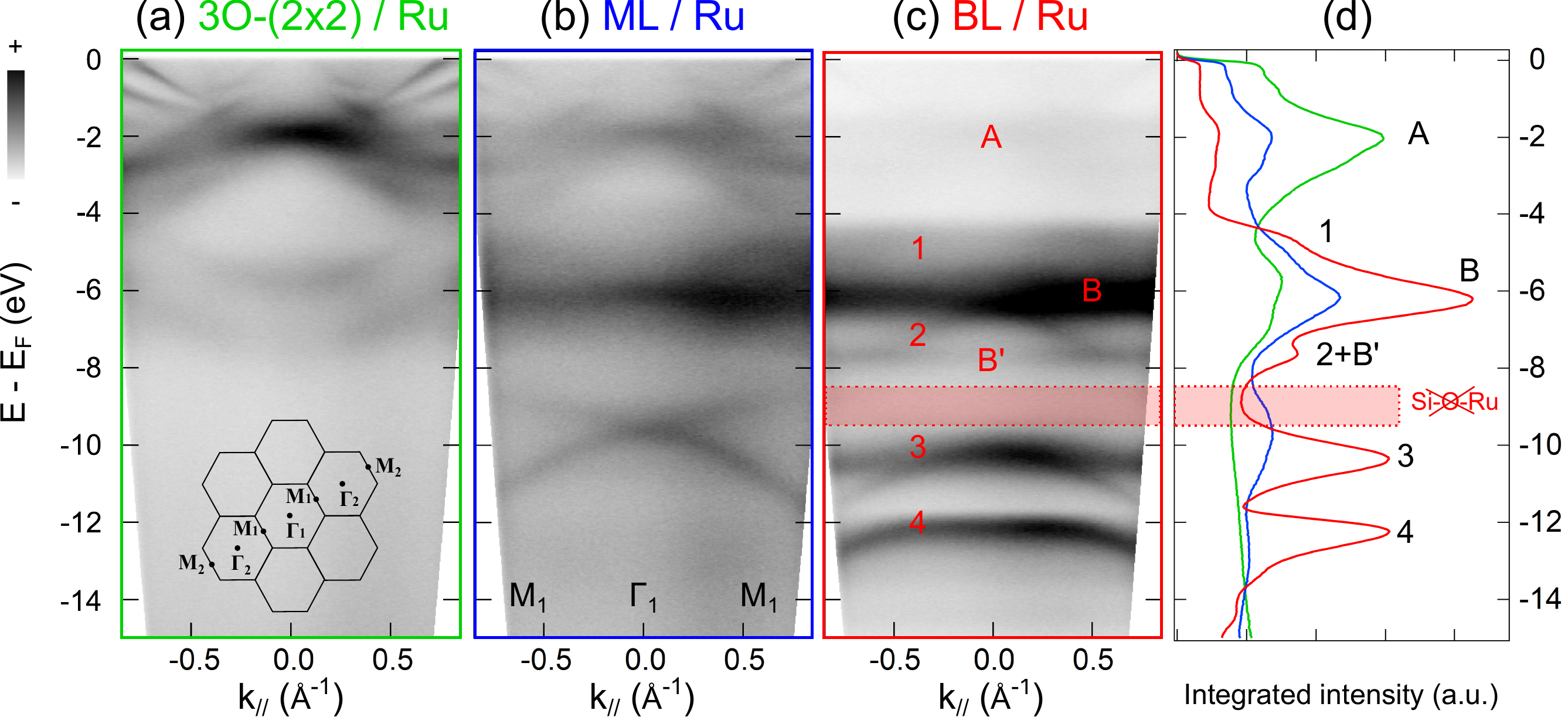} 
\end{center}
\caption{\label{Figure2} ARPES spectra using HeII radiation ($h \nu =  40.8$ eV) along the $M_{1}-\Gamma_{1}-M_{1}$ high-symmetry direction for (a) 3O--(2$\times$2)/Ru(0001), (b) ML silicon oxide and (c) crystalline BL silicon oxide. In the low part of panel (a), the (2$\times$2) surface BZ is represented. (d) Corresponding integrated density of states (IDOS) extracted from ARPES data. The red dashed box highlights the depletion of spectral weight observed in the BL case, related to the absence of Si--O--Ru contributions as observed for the ML \cite{kremer2019electronic}.}
\end{figure*}

\section*{Signatures of the formation of the bilayer silicon oxide}

 To get a better understanding of the formation of the BL on Ru(0001), we follow the evolution of the O 1s core level spectra with XPS from the \hbox{3O--(2$\times$2)/Ru(0001)} superstructure to the ML silicon oxide and finally to the BL silicon oxide (see \hbox{figure~\ref{Figure1}(d)}). The 3O--(2$\times$2)/Ru(0001) (green) spectrum exhibits only one contribution which corresponds to equivalent O atoms on Ru hollow sites (O--Ru). In the ML case (blue), we  resolve only three peaks. Nevertheless, as demonstrated in our previous work, the spectrum can be deconvoluted into four contributions corresponding to Si--O--Si (531.1~eV), \hbox{Si--O--Ru} (530.1~eV, left shoulder in the central peak, and 529.1~eV) and O--Ru bonds (529.9~eV, second contribution in the central peak) \cite{kremer2019electronic}, in agreement with a simulated XPS spectrum \cite{yang_thin_2012}. These different contributions take their origin from the existence of different kinds of bonds in the ML structure, as visible in the ball model in figure~\ref{Figure1}(c).

After growth of BL silicon oxide (red curve in figure~\ref{Figure1}(d)), the O 1s core level exhibits essentially two components, one at 529.9 eV BE corresponding to O bound on the hollow sites of Ru(0001) \hbox{(O--Ru)}, and another one, the prominent one, at a BE of \hbox{531.5 eV}. This component is reminiscent of the one occurring at 531.1 eV in the ML on Ru(0001), where it was ascribed to O atoms bound to two Si atoms, on the top oxide layer, as discussed above. Oxygen atoms with such a chemical environment are also present in BL silicon oxide (figure \ref{Figure1}(b)) and they are naturally those corresponding to the 531.5 eV component. Another kind of O atom, also surrounded by two Si atoms, is found in the median plane of BL silicon oxide. The BE of the corresponding O 1s core level is predicted to differ by only 0.13 eV from the one corresponding to O atoms in the top and bottom planes of silicon oxide \cite{wlodarczyk_tuning_2012}. We hence expect that the Si--O--Si peak centered at 531.5 eV actually consists of two components, which we cannot resolved due to the limited energy resolution of our experiment and the residual disorder in the system. Finally, the remarkable point is the absence of \hbox{Si--O--Ru} bonds, contrary to the ML case, confirming the disconnection of the BL silicon oxide from Ru(0001), in good agreement with the literature \cite{loffler_growth_2010,yang_thin_2012} and with what is expected from a van der Waals interface.

\section*{Observation of dispersive electronic states in bilayer silicon oxide} 

We now discuss the evolution of the band structure mapped by ARPES, recorded along the $M_{1}-\Gamma_{1}-M_{1}$ direction of the BZ (see figure~S2), for the 3O--(2$\times$2)/Ru(0001), the ML and the BL silicon oxides (figure \ref{Figure2}(a-c)). The corresponding integrated density of states (IDOS)  data are extracted from the momentum-resolved ARPES maps and shown in \hbox{figure \ref{Figure2}(d).}

The dispersions and IDOS of the filled electronic bands, measured after the growth of the BL silicon oxide, have only few features in common with the data measured for the 3O--(2$\times$2)/Ru(0001) and ML silicon oxide. First, a contribution centered close to -2~eV (label A) is found in all three systems. It is assigned to bonds between the substrate and individual O atoms, and its intensity is the weakest in the case of BL silicon oxide. This contribution seems very weak in the ARPES data (figure \ref{Figure2}(c)), and is better resolved in the IDOS spectrum (red curve in \hbox{figure \ref{Figure2}(d)).} This intensity reduction is a consequence of outgoing photoelectrons absorption within the material above the interfacial O atoms, which is the thickest in the case of BL silicon oxide. Second, ML and BL silicon oxide both exhibit a broad peak centered close to -6~eV (label B), whose line-shape is much different from the case of 3O--(2$\times$2)/Ru(0001). This peak was already observed in ultraviolet photoemission spectroscopy of BL silicon oxide on Ru(0001) \cite{wlodarczyk_tuning_2012}, and of silicon and germanium oxides without a metal substrate \cite{distefano_photoemission_1971,fischer_electronic_1977}, and it is related to \hbox{O 2p} non-bonding states in Si--O--Si bridges.

Beyond relative differences in the intensities of some of the bands, striking qualitative differences are observed between the band structure of the ML and BL silicon oxide. Noteworthy, a dip of intensity is observed around -9~eV for the BL, where the signature of \hbox{Si--O--Ru} bonds is expected (and observed) in the case of the ML \cite{kremer2019electronic}. This further confirms the decoupling between the BL and the substrate. In addition, a contribution is observed at -7.5~eV, and two intense bands are centered at -10.5~eV and -12.5~eV for the BL, while two weak bands  are centered at -9.5~eV and -13~eV for the ML.

\begin{figure*}
\centering
\includegraphics[scale=0.187]{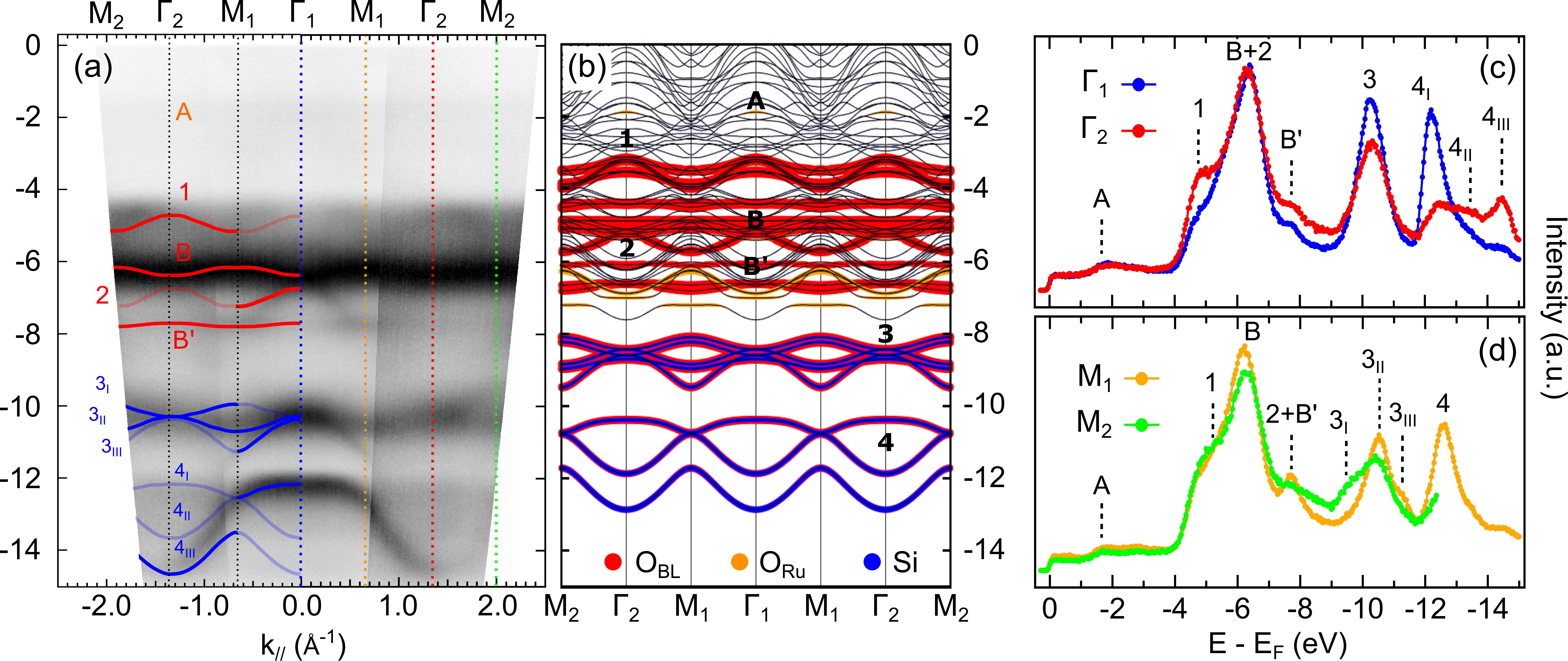}
\caption{\label{Figure3} (a) ARPES cut in the electronic band structure along the $M_{2}-\Gamma_{2}-M_{1}-\Gamma_{1}-M_{1}-\Gamma_{2}-M_{2}$ high-symmetry direction, measured using HeII radiation ($h \nu$ = 40.8 eV). Red and blue curves represent the main DFT calculated bands from (b), shifted by -1.7~eV. Thick and thin curves have strong and weak experimental spectral weight, respectively. (b) Electronic band structure calculated with DFT for crystalline BL silicon oxide. Black, red, orange and blue colors correspond respectively to Ru, O in the BL sheet, O connected to Ru, and Si character of the bands. EDCs taken at (c) the $\Gamma$ points and (d) the $M$ points of the BZ, corresponding to colored vertical dashed lines in the right part of panel (a). In the text, we associate the maximum of each contribution of these EDCs, at the different high symmetry points of the BZ ($\Gamma$ and $M$), to the location of the different electronic bands of the BL.}
\end{figure*}

The differences between the BL and the ML band structures are even more obvious in the extended BZ, as depicted in figure~\ref{Figure3}(a). To support the determination of the periodicity and dispersion of the bands inferred from the ARPES maps, and alleviate the locally difficult analysis of the latter data where signals are too weak due to photoemission matrix elements effects, energy distribution curves (EDCs) were also extracted at the $M$ and $\Gamma$ points of the first and second BZs (\hbox{figure \ref{Figure3}(c,d)).}

Four groups of more-or-less dispersing bands are specifically observed in the BL, in the ranges \hbox{[-4.9,-5.2]~eV} (band 1), \hbox{[-6.2,-7.8]~eV} (bands 2 and B'), \hbox{[-9.8,-11.2]~eV} (group of bands 3), and \hbox{[-12.1,-14.5]~eV} (group of bands 4). The band centered around -2~eV (band A), which stems from the individual O atoms on Ru(0001) and not from the silicon oxide, and the non-dispersive bands centered around -6~eV and -7.8~eV (bands B and B'), ascribed to non-bonding O states, will not be discussed further.

Band 1 is mainly visible in the second BZ, and spans from -4.9~eV ($\Gamma$ point) to -5.2~eV ($M$ point). Band 2 is only visible in the first BZ, spanning from -6.2~eV ($\Gamma$ point) to -7.8~eV ($M$ point).

The group of bands 3 consists of three dispersive bands, denoted 3$_\mathrm{I}$, 3$_\mathrm{II}$, and 3$_\mathrm{III}$. These bands are degenerate at the $\Gamma$ point, yielding a strong peak at -10.2~eV in the corresponding EDC (figure~\ref{Figure3}(c)). Band 3$_\mathrm{I}$ is mainly visible in the second BZ and spans from -9.8~eV ($M$ point) to -10.2~eV ($\Gamma$ point). The semi-flat band 3$_\mathrm{II}$ has spectral weight throughout the first and second BZs, and contributes to the peak located at -10.5~eV in the EDC at the $M$ points (figure~\ref{Figure3}(d)). Band 3$_\mathrm{III}$ has spectral weight in the first BZ only, and strongly disperses, spanning from -10.2~eV ($\Gamma$ point) to -11.2~eV ($M$ point).

The group of bands 4 also consists of three dispersive bands, denoted 4$_\mathrm{I}$, 4$_\mathrm{II}$, and 4$_\mathrm{III}$. Band 4$_\mathrm{I}$ spans from -12.1~eV ($\Gamma$ point) to -12.5~eV ($M$ point), and has stronger spectral weight in the first BZ. Band 4$_\mathrm{II}$ has globally low spectral weight and spans from -12.5~eV ($M$ point) to -13.5~eV ($\Gamma$ point). Band 4$_\mathrm{III}$ has spectral weight mostly in the second BZ, and spans from -12.5~eV ($M$ point) to -14.5~eV ($\Gamma$ point). Across the degeneracy point ($M$ point), the spectral weight seems partially transferred from band 4$_\mathrm{I}$ to band 4$_\mathrm{III}$. This is likely the signature of a small band gap, not resolved experimentally due to spectral broadening: the spectral weight is mainly located on the folded bands, \textit{i.e.} on band 4$_\mathrm{I}$ in the first BZ and band 4$_\mathrm{III}$ in the second BZ.

\section*{Origin of the dispersive bands} 

Figure~\ref{Figure3}(b) displays the band structure calculated for the BL silicon oxide with DFT along the $M-\Gamma-M$ high-symmetry direction, using the structural model depicted in figure~\ref{Figure1}(a,b). The bands are displayed with different colours, according to their Ru, O, and Si character. From the Fermi level E\textsubscript{F} down to -7.5~eV, numerous bands from the substrate are observed, together with a few bands involving individual O atoms (O--Ru) and others involving O atoms within the BL silicon oxide (Si--O--Si). Below -8~eV, the electronic bands exclusively stem from O and Si atoms inside the BL silicon oxide (Si--O--Si).

There is a strong resemblance between the calculated electronic band structure and the ARPES data: in the former, dispersive bands in the \hbox{[-3.0,-4.0]~eV}, \hbox{[-5.0,-5.5]~eV}, \hbox{[-8.0,-9.5]~eV}, and \hbox{[-10.3,-13.0]~eV} ranges are reminiscent of the groups of bands 1-4 in the latter. The calculated semi-flat bands in the \hbox{[-4.5,-6.5]~eV} range, corresponding to non-bonding states, seem to correspond to the experimental signal in the \hbox{[-5.0,-7.5]~eV} range. Altogether, the calculated band structure is globally shifted to higher energies, by about 1.7~eV. Besides, the calculation sometimes predict two-to-three close-by bands where the experimental data would point to a single band (1, 3$_\mathrm{I}$, 3$_\mathrm{II}$), presumably due to the finite experimental energy resolution. Note that due to the extreme surface sensitivity of ARPES, bands from the buried Ru and individual O atoms have minor contribution in the experiments.

\begin{figure*}
\centering
\includegraphics[scale=1.25]{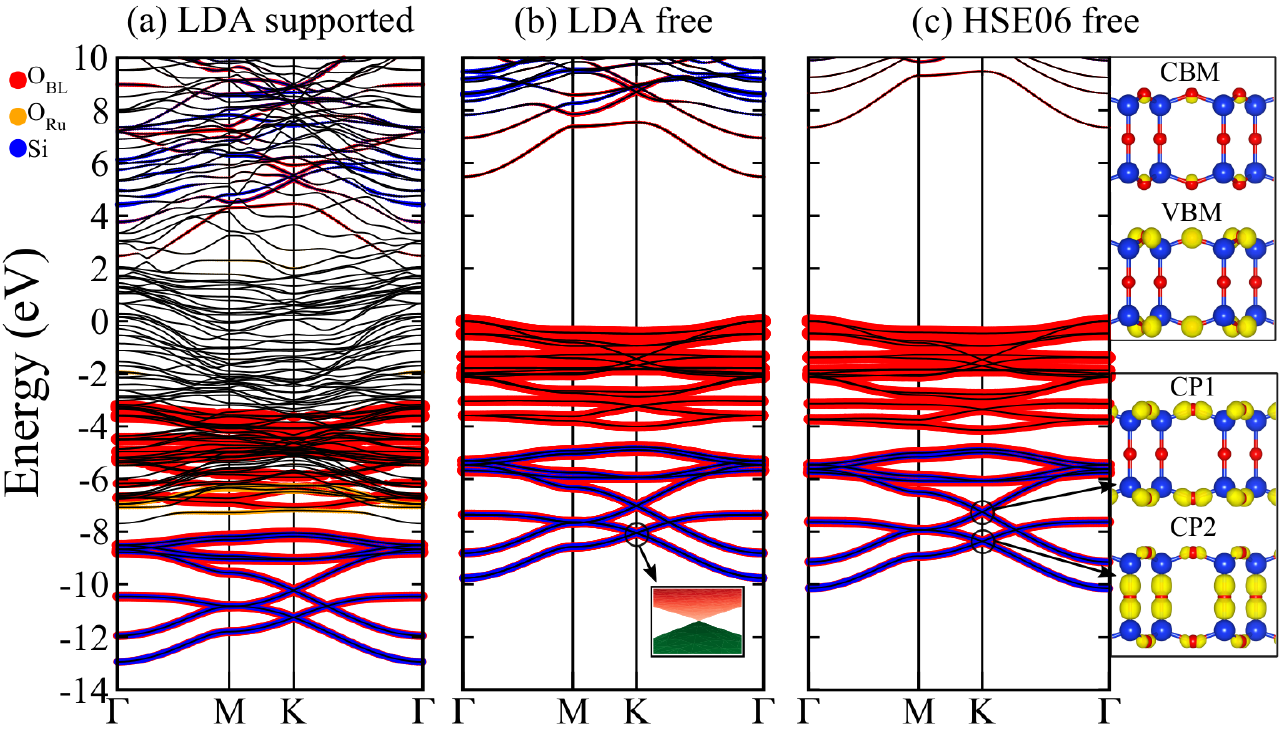}
\caption{\label{Figure4} Calculated band structure of BL silicon oxide (a) on Ru(0001), using the LDA framework, and in the absence of the substrate (free-standing) with (b) LDA and (c) HSE06 functionals. Right panels correspond to the partial charge densities of the (top) CBM and VBM orbitals at the $\Gamma$-point with an isosurface value of 0.012~$e$/{\AA}$^3$ (bottom) and linear dispersive bands  around the crossing $K$-point at -7.3 eV (CP1) and -8.41 eV (CP2) within an energy range of $\pm 0.07$ eV and isosurface value of 0.001~$e$/{\AA}$^3$. The inset in panel (b) corresponds to the 3D band structure calculation in the vicinity of the CP2 point, as detailed in figure~S5. Orange, red and blue dots correspond to the projection on the oxygen atom on top of Ru (O\textsubscript{Ru}), oxygen from the BL (O\textsubscript{BL}) and silicon (Si) atoms in the BL silicon oxide, respectively. Zero energy corresponds to (a) the Fermi level and (b),(c) the VBM.}

\end{figure*}

Summarizing partial density of states (PDOS) calculations presented in figure~S4, we  now discuss the orbital character of the group of bands 1-4. Band 1 is dominated by contributions from s+p\textsubscript{x}+p\textsubscript{y} and p\textsubscript{z} orbitals of O atoms involved in Si--O--Si bonds, both those in the outer planes of the BL and those in its median plane. In this sense the band has both in-plane and out-of-plane characters, unlike band 2, which is dominated by in-plane s+p\textsubscript{x}+p\textsubscript{y} O and Si orbitals \hbox{(Si--O--Si bonds)} in the two outer planes of the structure.

Similarly, bands 3\textsubscript{I} and 3\textsubscript{III} are dominated by (mostly) in-plane O and Si orbitals, within the Si--O--Si bonds in the two outer planes of the structure. Band 3\textsubscript{II}, on the contrary, involves out-of-plane p\textsubscript{z} orbitals of O atoms, and is thus related to the out-of-plane \hbox{Si--O--Si} bridges connecting the two outer planes of the structure.

Finally, bands 4\textsubscript{II} and 4\textsubscript{III} originate from orbitals involved in the in-plane Si--O--Si bonds of the two outer planes, and band 4\textsubscript{I} from orbitals in the out-of-plane \hbox{Si--O--Si bridges} between these two planes.

Interestingly, the most weakly dispersing bands, 3\textsubscript{II} and 4\textsubscript{I}, which are in fact semi-flat at least in extended wavevector ranges, are both related to the out-of-plane \hbox{Si--O--Si} bridges, \textit{i.e.} in a direction corresponding to strong confinement conditions ($z$) characteristic of a 2D material. More significant dispersions are systematically observed for those bands (1, 2, 3\textsubscript{I}, 3\textsubscript{III}, 4\textsubscript{II}, 4\textsubscript{III}) corresponding to in-plane delocalized electronic states, hosted by the extended network of mostly in-plane \hbox{Si--O--Si} bonds of the two outer planes of the 2D material.

\section*{Empty states and band gap of bilayer silicon oxide}

The ARPES data reveal that the as-grown BL silicon oxide has a VBM at the edge of band 1, \textit{i.e.} at the $\Gamma$ point, -4~eV below E\textsubscript{F} (electronic states found around -2~eV stem from individual O atoms on Ru(0001), not from the BL silicon oxide itself). In the DFT calculations, the VBM is up-shifted to -3.25~eV (figure~\ref{Figure4}(a)).

Our ARPES measurements give no information on empty electronic states, hence at this stage we can only assume that the conduction band minimum (CBM) is above (or at) E\textsubscript{F}. This yields a lower estimate band gap value of 4~eV, an estimate that we will later revise.

The DFT calculations, performed using the LDA functional, predict a CBM at the $\Gamma$ point, +2.47~eV above E\textsubscript{F} (figure~\ref{Figure4}(a)). In this framework BL silicon oxide is hence predicted to be a direct, 5.72~eV band gap semiconductor. In passing, we note that the cut chosen to compute the electronic band structure for figure~\ref{Figure4}(a) crosses the K point, where several linearly-dispersing bands cross, in the valence (-10~eV, \hbox{-11~eV}) and conduction (+5.5
~eV) bands. The linearity of these bands along all the directions of the BZ is demonstrated in figure~S5 and in the inset of figure~\ref{Figure4}(b) where we performed 3D band structure calculations in the vicinity of the K point. Similar linear dispersions and crossing have been observed in the ML silicon oxide \cite{kremer2019electronic} and, before that, in a variety of materials \cite{ohta2006controlling,wehling2014dirac,requist2015spin,bahramy2018ubiquitous,horio2018two}. We note that just like in graphene where similar linear dispersions were extensively studied, the lattice of BL silicon oxide is a honeycomb one. This suggests a possible massless Dirac fermion behaviour for electrons in specific electronic bands of this material as well (in contrast to graphene, well below the Fermi level, though), with a sub-lattice symmetry controlling the formation of the Dirac cones. Noteworthy is the presence of the two linear crossings in BL silicon oxide, shifted in energy one with respect to the other. This is reminiscent of what has been predicted in BL graphene with a particular kind of atomic stacking \cite{tabert} unlike the naturally occurring Bernal one. In this particular stacking, all C atoms in one layers sit directly on top of a C atom in the layer below. This kind of atomic stacking is precisely the one found between the top and bottom planes in BL silicon oxide, which stresses the BL character, from the electronic point of view, of the material. Additionnal arguments related to the topology of the system have to be brought forth to support the claim that these features are true Dirac cones.

Free-standing BL silicon oxide shows a remarkably similar electronic band structure (figure~\ref{Figure4}(b)). Obviously, in the absence of the numerous Ru states, this band structure is simpler, but features the same \hbox{Si--O--Si bands} as in supported (on Ru(0001)) BL silicon oxide. These bands are shifted to higher energies, almost rigidly, and the band gap is very similar, \textit{i.e.} direct and equal to 5.48~eV. The negligible influence of the Ru substrate on the band structure confirms its weak interactions with the BL silicon oxide. The weak interaction is further corroborated by the similar band structure obtained by taking into account van der Waals interactions, as detailed in figure~S6 and S7.


\begin{figure*}
\centering
\includegraphics[width=160mm]{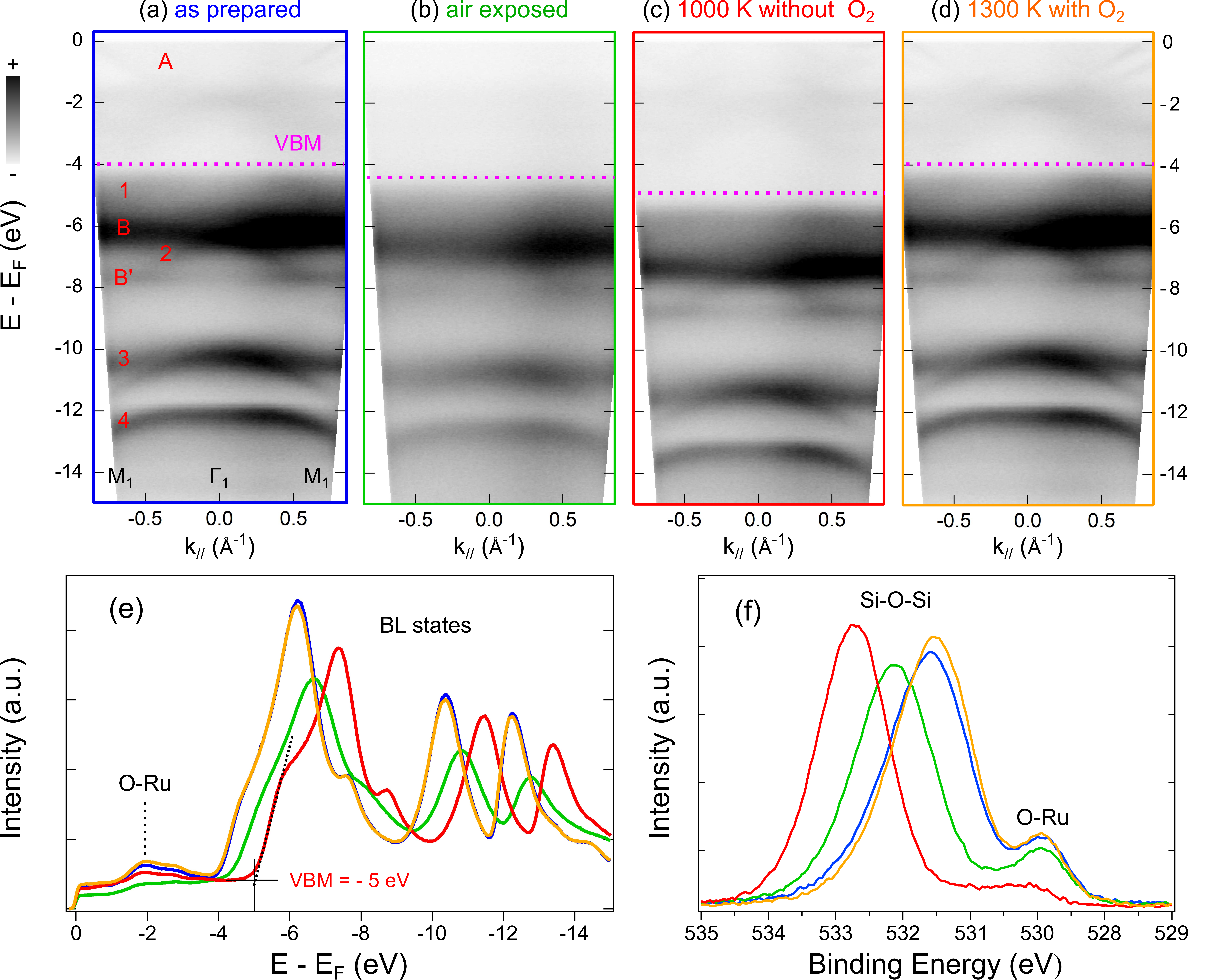}
\caption{\label{Figure5} ARPES cuts in the electronic band structure using HeII radiation ($h \nu$ = 40.8 eV) along the $M_{1}-\Gamma_{1}-M_{1}$ direction  for the BL silicon oxide (a) as prepared on Ru(0001), (b) after exposure to air, (c) annealed at $1000$\,K without O\textsubscript{2} and (d) annealed at $1300$\,K  with  O\textsubscript{2}. The VBM is highlighted by a horizontal pink dashed line. Corresponding (e) IDOS extracted from the ARPES data and (f) high-resolution XPS spectra of the O 1s core level recorded with a photon energy of 1486.6 eV.}
\end{figure*}


The structure of free-standing BL silicon oxide is simple enough to perform more advanced DFT calculations, using the HSE06 exchange correlation hybrid functional. For the kind of system of interest here, electronic band structures calculated with this functional better reproduce experimental data than with the LDA one, as shown in the case of bulk $\alpha$-quartz \cite{Alkauskas2007}, for which a 8.3~eV band gap was derived (\textit{versus} 5.8~eV with a semi-local functional), close to the experimental value of $\sim$ 9eV. In the case of BL silicon oxide, the HSE06 estimate of the band gap is of 7.36~eV (figure~\ref{Figure4}(c)), in agreement with previous hybrid calculations (7.2 eV) \cite{Gao2016}, a significantly larger estimate than within the LDA framework. Nevertheless, it is worth to mention that the LDA well describes the occupied part of the band structure. The most important modification is the description of the unoccupied states which are rigidly shifted to positive energy in the case of the HSE functional.

Using DFT calculations, we can also analyse the nature of the electronic states in real space. At the VBM and CBM, which are located at the $\Gamma$ point, essentially pure O states are found (\hbox{figure~\ref{Figure4}(a-c)} and figure S4). There, the dispersion is vanishing, and consistently the partial charge density appears localized on O atoms in the top and bottom planes of the BL (figure~\ref{Figure4}(c)). On the contrary, at the K point the dispersion is significant and the bands have mixed Si and O character (figure~\ref{Figure4}(a-c) and figure S4). The corresponding electronic states are more delocalised. In particular, the electronic density associated to the CP1 point is confined to the top and bottom planes of the structure, where it seems in-plane delocalised (figure~\ref{Figure4}(c)).


\section*{Tuning of the electronic properties by external parameters} 

In the following, we address the electronic structure modifications of the BL silicon oxide  induced upon exposure to air and as a function of the amount of individual O atoms directly bound to Ru(0001). In both cases, we discuss the electronic structure evolution by means of  ARPES (figure~\ref{Figure5}(a-e)) and XPS (figure~\ref{Figure5}(f)).

We first prepared a crystalline BL silicon oxide, showing a characteristic O 1s core level and a well defined band structure together with pronounced \hbox{O--Ru} states at -2 eV below E\textsubscript{F} (figure~\ref{Figure5}(a,e,f)). The BL VBM  is located -4 eV below E\textsubscript{F}. The position of the VBM is estimated as the intersection between the linearly extrapolated spectral leading edge and the baseline \cite{VBM} (as an example, see the red curve in figure~\ref{Figure5}(e)).

Then, we exposed the sample to air and introduced it back in ultra-high vacuum after four hours. The intensity of the O 1s core level is globally reduced due to surface contamination but the line shape of \hbox{Si--O--Si} and O--Ru contributions is not modified, indicating an intact chemical composition and binding configuration (figure~\ref{Figure5}(f)). The most striking effect is the shift at high BE of the Si--O--Si component and the unchanged position of the O--Ru peak that we will discuss at the end of this section. The BL band structure is still resolved even though bands are slightly broadened (figure~\ref{Figure5}(b,e)). Furthermore, the VBM is also shifted to high BE in comparison to the as-prepared BL, to -4.4 eV. Overall, the electronic properties of the BL silicon oxide are globally preserved upon exposure to air.\

Subsequently, the BL has been first annealed to $1000$\,K without O\textsubscript{2} and then to $1300$\,K under $3 \times 10^{-6}$\,mbar O\textsubscript{2}. After the first annealing, the \hbox{Si--O--Si peak} in the O 1s core level shifts to high BE in comparison to the as-prepared BL (+1.2 eV), while the O--Ru contribution is marginally shifted. The intensity of the former is constant but the intensity of the latter is drastically reduced.  After the high-temperature annealing with O\textsubscript{2}, the O 1s core level recovers its initial shape, \textit{i.e.} the \hbox{Si--O--Si} peak shifts back to 531.5~eV BE and the O--Ru peak recovers a lot of intensity. During the process, both intensity and full width at half maximum of the Si--O--Si contribution are unchanged, demonstrating an intact chemical composition for the BL sheet. The main difference between the two samples, annealed with or without O\textsubscript{2}, comes from the chemisorbed oxygen on the Ru surface.\

These features, and in particular these energy core level shifts, have already been observed by \hbox{W\l{}odarczyk \emph{et al.}} \cite{wlodarczyk_tuning_2012} and their origin has been explained by \hbox{Wang \emph{et al.}} \cite{wang2017energy}. By annealing with (without) oxygen it is possible to increase (decrease) the amount of O atoms at the metal-oxide interface. When this amount is low, the distance of the silicon oxide from the metal surface is relatively small and the influence of the electrons transferred from the insulating sheet to the Ru surface dominates: a positive interface dipole is formed. The effect is the diminution of the work function and by consequence the shift to high BE of the related Si--O--Si contribution in the \hbox{O 1s} core level. The positions of the Si--O--Si and O--Ru contributions are separated by 2.75 eV. Extrapolating the data from \hbox{Wang \emph{et al.}}  \cite{wang2017energy}, we conclude that after annealing at $1000$\,K without O\textsubscript{2}, about 0.5 individual O atom directly binds to Ru(0001) per BL silicon oxide unit cell (as defined in figure~\ref{Figure1}(a), \textit{i.e.} half the size of the rectangular one defined by \hbox{Wang \emph{et al.}} \cite{wang2017energy}).

When the density of O atoms on the Ru surface increases, the height of the BL silicon oxide increases, and as a consequence, the number of transferred electrons from the BL to the surface decreases \cite{wang2017energy}. This promotes charge transfer from the Ru substrate to chemisorbed O, gives rise to a negative surface dipole and to an associated increase of the work function, leading to a core level shift to low BE this time. In our case, after annealing with oxygen, the Si--O--Si peak is located at 531.5~eV, \textit{i.e.} 1.60~eV appart from the O--Ru contribution. Based on the analysis introduced by Wang \emph{et al.} \cite{wang2017energy}, we deduce that about 3 individual O atoms now bind to Ru(0001) per BL silicon oxide unit cell. The two effects mentioned above (interface and surface charge transfers) are competing, and their balance depends on the number of chemisorbed oxygen atoms at the surface.

The energy shift of the Si--O--Si and intensity variation of the O--Ru core levels overall witness a reversible tuning of the electronic properties of the BL induced by a varying oxygen concentration at the metal-oxide interface. Additional arguments to support this scenario are given by the analysis of the Ru 3d\textsubscript{5/2} core level. The Ru 3d\textsubscript{5/2} core level is  very sensitive to the surface oxidation \cite{kremer2019electronic,lizzit_surface_2001}. Data measured on the BL silicon oxide, the \hbox{3O--(2$\times$2)/Ru(0001)} and the pristine Ru(0001) surfaces are presented in figure~S8. The larger (fewer) the number of chemisorbed O atoms, the more (less) intense is the O--Ru contribution at 281 eV BE,  and the less (more) intense are the Ru bulk and surface ones at \hbox{280 eV} and 279.8 eV BE respectively. In figure~S8(a), it is obvious that the O--Ru contribution is enhanced when the BL is annealed with oxygen (orange curve) while it is reduced in favor to bulk and surface contributions when the sample is annealed without oxygen (red curve). Note that the O--Ru and Ru surface contributions in  figure~S8(c) are better resolved in comparison to figure~S8(b), due to both the better energy resolution (70 meV with synchrotron radiation in comparison to 300 meV with a laboratory X-rays source) and the higher surface sensitivity at h\hbox{$\nu$ = 350 eV}.

We can now rationalise the ARPES observations. Figure~\ref{Figure5}(c,d) shows the spectra after annealing without and with oxygen respectively. As we have just seen, these two measurements correspond to the band structures in the presence of more or less O atoms on the Ru(0001) surface. Strikingly, in both cases, we recover the characteristic bands of the highly crystalline BL discussed above, in particular the group of bands \hbox{1-4}. We observe two main differences. First, the VBM after annealing without and with oxygen are located at -5~eV and -4~eV below E\textsubscript{F} respectively. This is also visible in the IDOS where all the spectroscopic signatures of the BL are shifted to high BE in the latter case. This difference has exactly the same origin as the core level shift observed in XPS for the \hbox{Si--O--Si} contribution in the O 1s core level. Additionally, the spectral weight of the O--Ru band at -2 eV below E\textsubscript{F} is much more intense in the case of the annealing with a partial pressure of oxygen. Its energy is not modified such as the O--Ru contribution in XPS. VBM shift and spectral weight variation of the O--Ru band hence appear, as we expect, directly proportionnal to the number of chemisorbed oxygen atoms on the Ru surface. Since the electronic band structure of the BL silicon oxide is not modified by the annealing without oxygen, but only rigidly shifted by 1~eV down to high BE, we can now estimate to 5 eV the lower bound of its band gap.

Finally, we briefly come back to the air exposure of the BL.  According to the O 1s core level in \hbox{figure \ref{Figure5}(f)}, and taking into account the surface contamination, one can say that the O--Ru contribution (and by consequence the number of oxygen at the interface) is not altered.  Nevertheless, we observe that the VBM and the \hbox{Si--O--Si} contribution in XPS are shifted to high BE in comparison to the fully oxidised case. It is likely due to the intercalation of contaminants at the interface during the atmosphere exposure, which could modify the surface dipole and then the associated work function.


\section*{Summary and concluding remarks}
We exploit the crystalline quality of epitaxial BL silicon oxide, grown on Ru(0001), to experimentally unveil its electronic band structure below E\textsubscript{F}. The material exhibits a series of well-defined dispersing and semi-flat electronic bands. Using first principle calculations we are able to determine the character of most of these bands and to highlight the existence of a set of two linear dispersing states at the K point consistent with the BL nature of the material. Some of the bands, which are strongly dispersing, involve in-plane hybridization between O and Si orbitals in the top and bottom layers of the structure. The semi-flat bands essentially originate from the Si--O--Si bridges between these bottom and top layers. We directly confirm the predicted high-band gap semiconductor character of the material. The VBM is indeed located between \hbox{-4} and -5 eV below E\textsubscript{F} depending on the amount of O atoms that can be chemisorbed onto the substrate (and underneath silicon oxide). This provides a lower bound estimate of the band gap (5~eV) in the material, which, according to our DFT calculations, is predicted to be direct and to amount 7.36 eV using the HSE exchange correlation functional which is known to well describe quasi-2D materials \cite{das2019band}.

Our work opens a route to the study of the largely unexplored electronic and optical properties of BL silicon oxide. Understanding whether the strong screening of Coulombic interactions due to the presence of the substrate can renormalise the electronic band structure and the band gap, compared to the free-standing material, seems for instance an important question to address. Such effects have been discussed in the context of graphene nanoribbons \cite{Kharche2016} and transition metal dichalcogenides \cite{Bruix2016,Waldecker2019}, and we propose controlled intercalation as a strategy to tune the dielectric environment of silicon oxide. A next step could focus on the exploration of the excitonic properties of the material, which we anticipate could be relevant for deep UV applications. To this respect, our DFT analysis of the character of the electronic bands involved at the CBM and VBM ($\Gamma$ point), and at other specific locations in reciprocal space, is a starting point for advanced optical studies, which could be based on photon spectroscopies (absorption, reflectance, photoluminescence) and \textit{ab initio} GW and Bethe-Salpeter equation calculations for instance. Such studies might unveil that BL silicon oxide hosts (especially when exfoliated from its substrate) a variety of excitonic complexes which, due to the minimal screening of Coulombic interactions in 2D, are expected to exhibit large BE. Whether BL silicon oxide can become a competitive alternative to hexagonal boron nitride for optoelectronics in this range is a totally open question at this stage. A second potential direction of research could relate to the implementation of BL silicon oxide into 2D van der Waals heterostructures. Tunnel junctions, with the BL silicon oxide as a tunnel barrier placed between two (semi)metallic or semiconducting 2D materials, may perform differently than hexagonal boron nitride, due to its larger thickness and the distinct nature of the electronic orbitals involved in the tunneling process -- there again, our DFT calculations might provide a valuable starting point for understanding the observed behaviours. The interactions between 2D materials in heterostructures being material-specific, we finally expect specific cross-talking effects between BL silicon oxide and other 2D materials, and to this respect too, knowledge about the nature of the surface electronic orbitals, derived from the DFT calculations, should be insightful.\\


\section*{Acknowledgments}
This work was supported by the 2DTransformers project under the OH-RISQUE program of the French National Research Agency (ANR-14-OHRI-0004). We would like to thank the team Da$\nu$m for the helpful assistance during the connection and the installation of the new SR--ARPES setup on the Tube.  G. K. acknowledges financial support from the Swiss National Science Foundation (SNSF) Grant No. P00P2$\_$170597. The DFT calculations were done using French supercomputers (GENCI, \# 6194) and the Predictive Simulation Center facility that gathers in Grenoble SPINTEC, L$\_$Sim and Leti. We thanks Professor N. Mousseau for useful discussions. C. G. acknowledges finantial support from the Community of Madrid through the project NANOMAGCOST CM-S2018/NMT-4321.


\section*{References}
\bibliography{biblio}

\begin{thebibliography}{10}

\bibitem{loffler_growth_2010}
Daniel L{\"o}ffler, John~J Uhlrich, M~Baron, Bing Yang, Xin Yu, Leonid
  Lichtenstein, Lars Heinke, Christin B{\"u}chner, Markus Heyde, Shamil
  Shaikhutdinov, et~al.
\newblock Growth and structure of crystalline silica sheet on
  \textsc{R}u(0001).
\newblock {\em Phys. Rev. Lett.}, 105(14):146104, 2010.

\bibitem{Huang2012}
Pinshane~Y Huang, Simon Kurasch, Anchal Srivastava, Viera Skakalova, Jani
  Kotakoski, Arkady~V Krasheninnikov, Robert Hovden, Qingyun Mao, Jannik~C
  Meyer, Jurgen Smet, et~al.
\newblock Direct imaging of a two-dimensional silica glass on graphene.
\newblock {\em Nano Lett.}, 12(2):1081--1086, 2012.

\bibitem{klemm_preparation_2016}
HW~Klemm, Gina Peschel, Ewa Madej, Alexander Fuhrich, Martin Timm, Dietrich
  Menzel, Th~Schmidt, and H-J Freund.
\newblock Preparation of silica films on \textsc{R}u(0001): A
  \textsc{LEEM}/\textsc{PEEM} study.
\newblock {\em Surf. Sci.}, 643:45--51, 2016.

\bibitem{kuhness2018two}
David Kuhness, Hyun~Jin Yang, Hagen~W Klemm, Mauricio Prieto, Gina Peschel,
  Alexander Fuhrich, Dietrich Menzel, Thomas Schmidt, Xin Yu, Shamil
  Shaikhutdinov, et~al.
\newblock A two-dimensional ‘zigzag’ silica polymorph on a metal support.
\newblock {\em J. Am. Chem. Soc.}, 140(19):6164--6168, 2018.

\bibitem{wlodarczyk_tuning_2012}
Rados{\l}aw W{\l}odarczyk, Marek Sierka, Joachim Sauer, Daniel L{\"o}ffler,
  JJ~Uhlrich, Xin Yu, Bing Yang, IMN Groot, S~Shaikhutdinov, and H-J Freund.
\newblock Tuning the electronic structure of ultrathin crystalline silica films
  on \textsc{R}u(0001).
\newblock {\em Phys. Rev. B}, 85(8):085403, 2012.

\bibitem{transfert}
Christin B{\"u}chner, Zhu-Jun Wang, Kristen~M Burson, Marc-Georg Willinger,
  Markus Heyde, Robert Schl{\"o}gl, and Hans-Joachim Freund.
\newblock A large-area transferable wide band gap 2d silicon dioxide layer.
\newblock {\em ACS Nano}, 10(8):7982--7989, 2016.

\bibitem{Gao2016}
Enlai Gao, Bo~Xie, and Zhiping Xu.
\newblock Two-dimensional silica: Structural, mechanical properties, and
  strain-induced band gap tuning.
\newblock {\em J. Appl. Phys.}, 119(1):014301, 2016.

\bibitem{Elias2019}
Christine Elias, Pierre Valvin, Thomas Pelini, A~Summerfield, CJ~Mellor,
  TS~Cheng, L~Eaves, CT~Foxon, PH~Beton, SV~Novikov, et~al.
\newblock Direct band-gap crossover in epitaxial monolayer boron nitride.
\newblock {\em Nat. Commun.}, 10(1):1--7, 2019.

\bibitem{Fang2014}
Hui Fang, Corsin Battaglia, Carlo Carraro, Slavomir Nemsak, Burak Ozdol,
  Jeong~Seuk Kang, Hans~A Bechtel, Sujay~B Desai, Florian Kronast, Ahmet~A
  Unal, et~al.
\newblock Strong interlayer coupling in van der waals heterostructures built
  from single-layer chalcogenides.
\newblock {\em Proc. Natl. Acad. Sci.}, 111(17):6198--6202, 2014.

\bibitem{Goratchev2012}
RV~Gorbachev, AK~Geim, MI~Katsnelson, KS~Novoselov, T~Tudorovskiy,
  IV~Grigorieva, Allan~H MacDonald, SV~Morozov, K~Watanabe, T~Taniguchi, et~al.
\newblock Strong coulomb drag and broken symmetry in double-layer graphene.
\newblock {\em Nat. Phys.}, 8(12):896--901, 2012.

\bibitem{Xian2019}
Lede Xian, Dante~M Kennes, Nicolas Tancogne-Dejean, Massimo Altarelli, and
  Angel Rubio.
\newblock Multiflat bands and strong correlations in twisted bilayer boron
  nitride: Doping-induced correlated insulator and superconductor.
\newblock {\em Nano Lett.}, 19(8):4934--4940, 2019.

\bibitem{kim_structural_1998}
Young~Dok Kim, Stefan Wendt, Stefan Schwegmann, Herbert Over, and Gerhard Ertl.
\newblock Structural analyses of the pure and cesiated
  \textsc{R}u(0001)--(2$\times$ 2)-3\textsc{O} phase.
\newblock {\em Surf. Sci.}, 418(1):267--272, 1998.

\bibitem{lichtenstein_atomic_2012}
Leonid Lichtenstein, Christin B{\"u}chner, Bing Yang, Shamil Shaikhutdinov,
  Markus Heyde, Marek Sierka, Rados{\l}aw W{\l}odarczyk, Joachim Sauer, and
  Hans-Joachim Freund.
\newblock The atomic structure of a metal-supported vitreous thin silica film.
\newblock {\em Angew. Chem. Int. Ed.}, 51(2):404--407, 2012.

\bibitem{yang_thin_2012}
Bing Yang, William~E Kaden, Xin Yu, Jorge~Anibal Boscoboinik, Yulia Martynova,
  Leonid Lichtenstein, Markus Heyde, Martin Sterrer, Rados{\l}aw W{\l}odarczyk,
  Marek Sierka, et~al.
\newblock Thin silica films on \textsc{R}u(0001): monolayer, bilayer and
  three-dimensional networks of [\textsc{S}i\textsc{O}$_{4}$] tetrahedra.
\newblock {\em Phys. Chem. Chem. Phys.}, 14(32):11344--11351, 2012.

\bibitem{lichtenstein2012probing}
Leonid Lichtenstein, Markus Heyde, Stefan Ulrich, Niklas Nilius, and
  Hans-Joachim Freund.
\newblock Probing the properties of metal--oxide interfaces: silica films on
  \textsc{M}o and \textsc{R}u supports.
\newblock {\em J. Phys. Condes. Mat.}, 24(35):354010, 2012.

\bibitem{dft-hk}
Pierre Hohenberg and Walter Kohn.
\newblock Inhomogeneous electron gas.
\newblock {\em Phys. Rev.}, 136(3B):B864, 1964.

\bibitem{dft-ks}
Walter Kohn and Lu~Jeu Sham.
\newblock Self-consistent equations including exchange and correlation effects.
\newblock {\em Phys. Rev.}, 140(4A):A1133, 1965.

\bibitem{Blochl1994}
Peter~E Bl{\"o}chl.
\newblock Projector augmented-wave method.
\newblock {\em Phys. Rev. B}, 50(24):17953, 1994.

\bibitem{Kresse1996}
Georg Kresse and J{\"u}rgen Furthm{\"u}ller.
\newblock Efficient iterative schemes for ab initio total-energy calculations
  using a plane-wave basis set.
\newblock {\em Phys. Rev. B}, 54(16):11169, 1996.

\bibitem{Ceperley1980LDA}
David~M Ceperley and Berni~J Alder.
\newblock Ground state of the electron gas by a stochastic method.
\newblock {\em Phys. Rev. Lett.}, 45(7):566, 1980.

\bibitem{Krukau2006HSE06}
Aliaksandr~V Krukau, Oleg~A Vydrov, Artur~F Izmaylov, and Gustavo~E Scuseria.
\newblock Influence of the exchange screening parameter on the performance of
  screened hybrid functionals.
\newblock {\em J. Chem. Phys.}, 125(22):224106, 2006.

\bibitem{grimme2006semiempirical}
Stefan Grimme.
\newblock Semiempirical \textsc{GGA}-type density functional constructed with a
  long-range dispersion correction.
\newblock {\em J. Comput. Chem.}, 27(15):1787--1799, 2006.

\bibitem{fischer_ultrathin_2015}
Frank~Daniel Fischer, Joachim Sauer, Xin Yu, Jorge~Anibal Boscoboinik, Shamil
  Shaikhutdinov, and Hans-Joachim Freund.
\newblock Ultrathin \textsc{T}i-\textsc{S}ilicate film on a \textsc{R}u(0001)
  surface.
\newblock {\em J. Phys. Chem. C}, 119(27):15443--15448, 2015.

\bibitem{kremer2019electronic}
Geoffroy Kremer, Juan~Camilo Alvarez~Quiceno, Simone Lisi, Thomas Pierron,
  C{\'e}sar Gonz{\'a}lez, Muriel Sicot, Bertrand Kierren, Daniel Malterre,
  Julien~E Rault, Patrick Le~F{\`e}vre, et~al.
\newblock Electronic band structure of ultimately thin silicon oxide on
  \textsc{R}u(0001).
\newblock {\em ACS Nano}, 13(4):4720--4730, 2019.

\bibitem{distefano_photoemission_1971}
TH~DiStefano and DE~Eastman.
\newblock Photoemission measurements of the valence levels of amorphous
  \textsc{S}i\textsc{O}$_{2}$.
\newblock {\em Phys. Rev. Lett.}, 27(23):1560, 1971.

\bibitem{fischer_electronic_1977}
B~Fischer, RA~Pollak, TH~DiStefano, and WD~Grobman.
\newblock Electronic structure of \textsc{S}i\textsc{O}$_{2}$,
  \textsc{S}i$_{x}$\textsc{G}e$_{1-x}$\textsc{O}$_{2}$, and
  \textsc{G}e\textsc{O}$_{2}$ from photoemission spectroscopy.
\newblock {\em Phys. Rev. B}, 15(6):3193, 1977.

\bibitem{ohta2006controlling}
Taisuke Ohta, Aaron Bostwick, Thomas Seyller, Karsten Horn, and Eli Rotenberg.
\newblock Controlling the electronic structure of bilayer graphene.
\newblock {\em Science}, 313(5789):951--954, 2006.

\bibitem{wehling2014dirac}
TO~Wehling, Annica~M Black-Schaffer, and Alexander~V Balatsky.
\newblock Dirac materials.
\newblock {\em Adv. Phys.}, 63(1):1--76, 2014.

\bibitem{requist2015spin}
Ryan Requist, Polina~M Sheverdyaeva, Paolo Moras, Sanjoy~K Mahatha, Carlo
  Carbone, and Erio Tosatti.
\newblock Spin-orbit interaction and dirac cones in d-orbital noble metal
  surface states.
\newblock {\em Phys. Rev. B}, 91(4):045432, 2015.

\bibitem{bahramy2018ubiquitous}
MS~Bahramy, OJ~Clark, B-J Yang, J~Feng, L~Bawden, JM~Riley, I~Markovi{\'c},
  F~Mazzola, V~Sunko, D~Biswas, et~al.
\newblock Ubiquitous formation of bulk \textsc{D}irac cones and topological
  surface states from a single orbital manifold in transition-metal
  dichalcogenides.
\newblock {\em Nat. Mater.}, 17(1):21--28, 2018.

\bibitem{horio2018two}
M~Horio, CE~Matt, K~Kramer, D~Sutter, AM~Cook, Yasmine Sassa, K~Hauser, Martin
  M{\aa}nsson, NC~Plumb, M~Shi, et~al.
\newblock Two-dimensional type-\textsc{II} \textsc{D}irac fermions in layered
  oxides.
\newblock {\em Nat. Commun.}, 9(1):1--7, 2018.

\bibitem{tabert}
C.~J. Tabert and E.~J. Nicol.
\newblock Dynamical conductivity of \textsc{AA}-stacked bilayer graphene.
\newblock {\em Phys. Rev. B}, 86:075439, 2012.

\bibitem{Alkauskas2007}
Audrius Alkauskas and Alfredo Pasquarello.
\newblock Effect of improved band-gap description in density functional theory
  on defect energy levels in $\alpha$-quartz.
\newblock {\em Physica B: Condensed Matter}, 401:670--673, 2007.

\bibitem{VBM}
Ahmad~Dauod Katnani and G~Margaritondo.
\newblock Microscopic study of semiconductor heterojunctions: photoemission
  measurement of the valance-band discontinuity and of the potential barriers.
\newblock {\em Phys. Rev. B}, 28(4):1944, 1983.

\bibitem{wang2017energy}
Mengen Wang, Jian-Qiang Zhong, John Kestell, Iradwikanari Waluyo, Dario~J
  Stacchiola, J~Anibal Boscoboinik, and Deyu Lu.
\newblock Energy level shifts at the silica/\textsc{R}u(0001) heterojunction
  driven by surface and interface dipoles.
\newblock {\em Top. Catal.}, 60(6-7):481--491, 2017.

\bibitem{lizzit_surface_2001}
S~Lizzit, Alessandro Baraldi, A~Groso, Karsten Reuter, Maria~Veronica
  Ganduglia-Pirovano, Catherine Stampfl, Matthias Scheffler, M~Stichler,
  C~Keller, W~Wurth, et~al.
\newblock Surface core-level shifts of clean and oxygen-covered
  \textsc{R}u(0001).
\newblock {\em Phys. Rev. B}, 63(20):205419, 2001.

\bibitem{das2019band}
Tilak Das, Giovanni Di~Liberto, Sergio Tosoni, and Gianfranco Pacchioni.
\newblock Band gap of 3\textsc{D} metal oxides and quasi-2\textsc{D} materials
  from hybrid density functional theory: \textsc{A}re dielectric-dependent
  functionals superior?
\newblock {\em J. Chem. Theory Comput.}, 15(11):6294--6312, 2019.

\bibitem{Kharche2016}
Neerav Kharche and Vincent Meunier.
\newblock Width and crystal orientation dependent band gap renormalization in
  substrate-supported graphene nanoribbons.
\newblock {\em J. Phys. Chem. Lett.}, 7(8):1526--1533, 2016.

\bibitem{Bruix2016}
Albert Bruix, Jill~A Miwa, Nadine Hauptmann, Daniel Wegner, S{\o}ren Ulstrup,
  Signe~S Gr{\o}nborg, Charlotte~E Sanders, Maciej Dendzik,
  Antonija~Grubi{\v{s}}i{\'c} {\v{C}}abo, Marco Bianchi, et~al.
\newblock Single-layer \textsc{M}o\textsc{S}$_{2}$ on \textsc{A}u(111):
  \textsc{B}and gap renormalization and substrate interaction.
\newblock {\em Phys. Rev. B}, 93(16):165422, 2016.

\bibitem{Waldecker2019}
Lutz Waldecker, Archana Raja, Malte R{\"o}sner, Christina Steinke, Aaron
  Bostwick, Roland~J Koch, Chris Jozwiak, Takashi Taniguchi, Kenji Watanabe,
  Eli Rotenberg, et~al.
\newblock Rigid band shifts in two-dimensional semiconductors through external
  dielectric screening.
\newblock {\em Phys. Rev. Lett.}, 123(20):206403, 2019.

\end{thebibliography}

\begin{center}



\section*{Supplementary material}

# Supplementary information for : dispersing and semi-flat bands in the wide band gap two-dimensional semiconductor bilayer silicon oxide

Geoffroy Kremer,*,†,‡ Juan Camilo Alvarez-Quiceno,¶ Thomas Pierron,‡ César González,§,∥ Muriel Sicot,‡ Bertrand Kierren,‡ Luc Moreau,‡ Julien E. Rault,⊥ Patrick Le Fèvre,⊥ François Bertran,⊥ Yannick J. Dappe,# Johann Coraux,@ Pascal Pochet,¶ and Yannick Fagot-Revurat‡

†Département de Physique and Fribourg Center for Nanomaterials, Université de Fribourg, CH-1700 Fribourg, Switzerland
‡Institut Jean Lamour, UMR 7198, CNRS-Université de Lorraine, Campus ARTEM, 2 allée André Guinier, BP 50840, 54011 Nancy, France
¶Department of Physics, IriG, Univ. Grenoble Alpes & CEA 38054 Grenoble France
§Departamento de Física de Materiales, Universidad Complutense de Madrid, E-28040 Madrid, Spain
∥Instituto de Magnetismo Aplicado UCM-ADIF, E-28230 Madrid, Spain
⊥Synchrotron SOLEIL, Saint-Aubin, BP 48, F-91192 Gif-sur-Yvette Cedex, France
#SPEC, CEA, CNRS, Université Paris-Saclay, CEA Saclay, 91191 Gif-sur-Yvette Cedex, France
@Univ. Grenoble Alpes, CNRS, Grenoble INP, Institut Néel, 38000 Grenoble, France

E-mail: geoffroy.kremer@unifr.ch

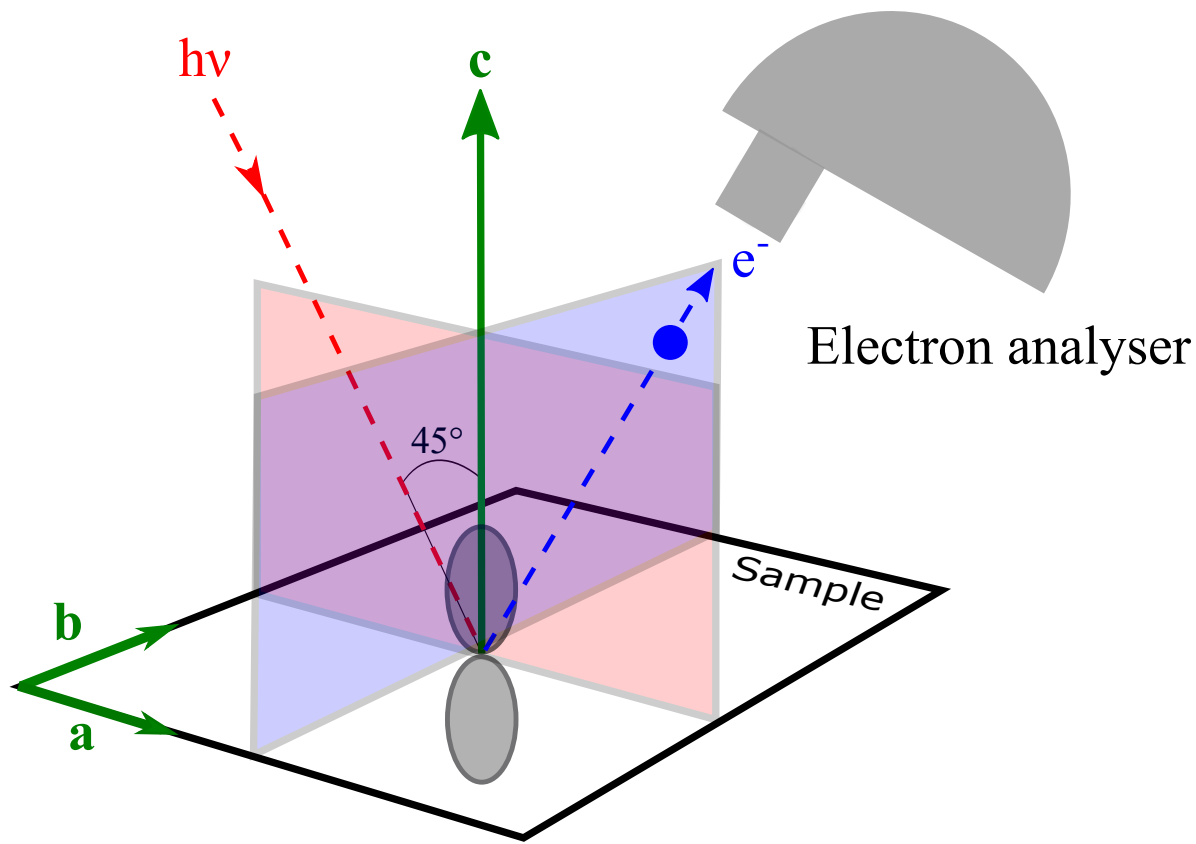

Figure S1: Experimental geometry of the photoemission experiment. Red and blue planes correspond respectively to the incidence plane of photons and to the detection plane of the analyser using vertical slit.

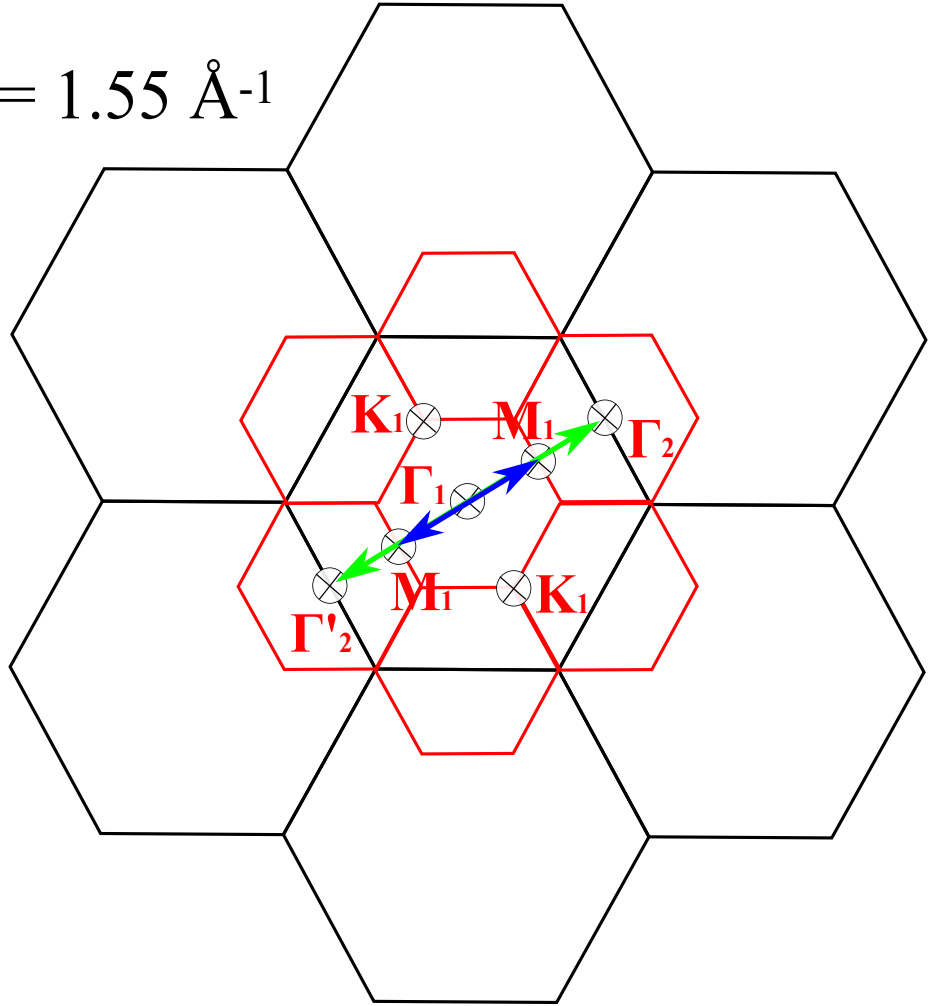

Figure S2: Representation of the extended Brillouin zones (BZ) of the (1×1) of Ru(0001) (black) and (2×2) of BL silicon oxide (red). The high symmetry points are given. Blue and green lines correspond to scan directions in figure S3, figure 2 and figure 3, respectively.

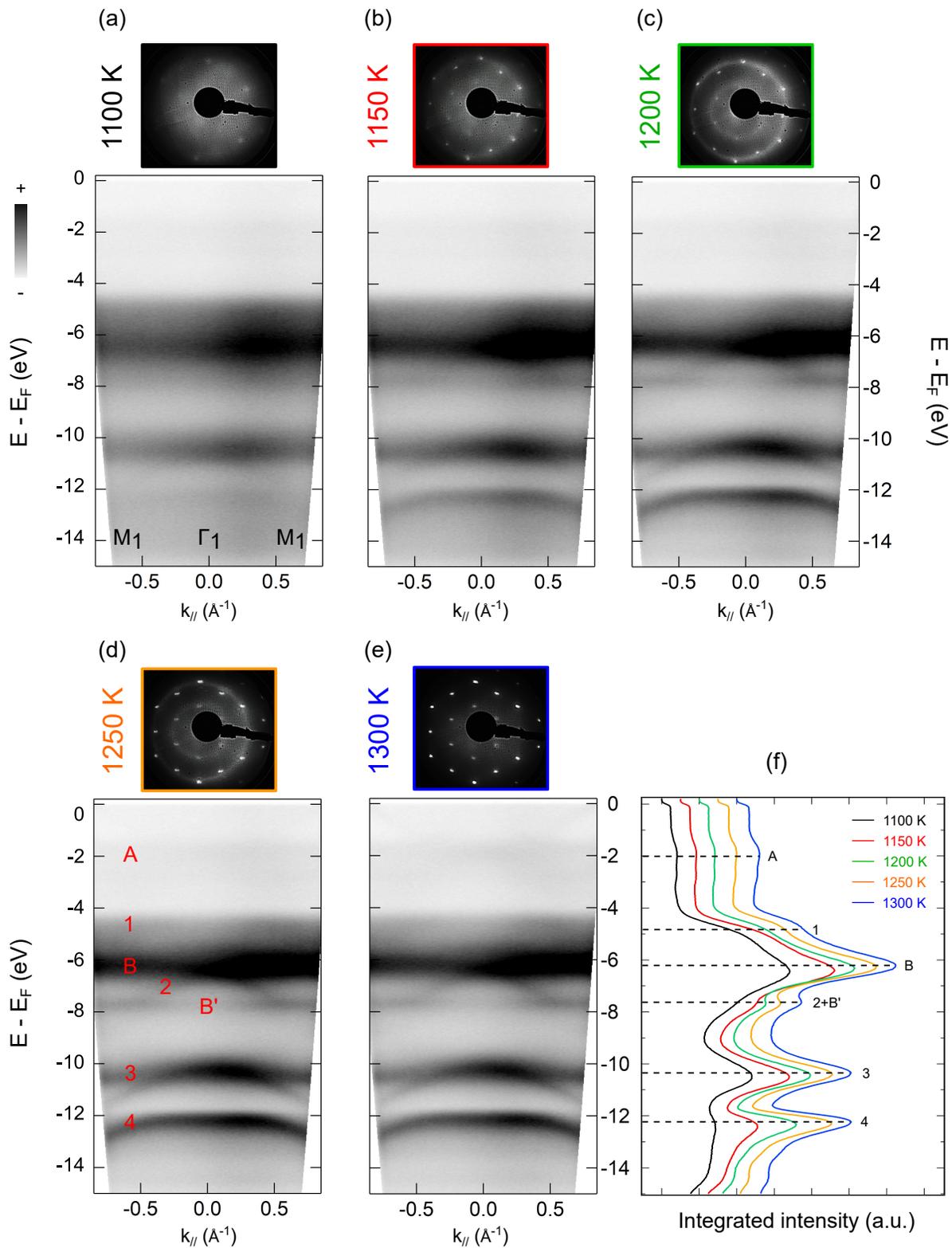

Figure S3: LEED patterns (top) and ARPES spectra (bottom) using HeII radiation ($h\nu = 40.8$ eV) along the $M_1 - \Gamma_1 - M_1$ direction for the BL silicon oxide as a function of the final annealing temperature (a) 1100 K, (b) 1150 K, (c) 1200 K, (d) 1250 K and (e) 1300 K. (f) Corresponding integrated density of states extracted from ARPES data.

4

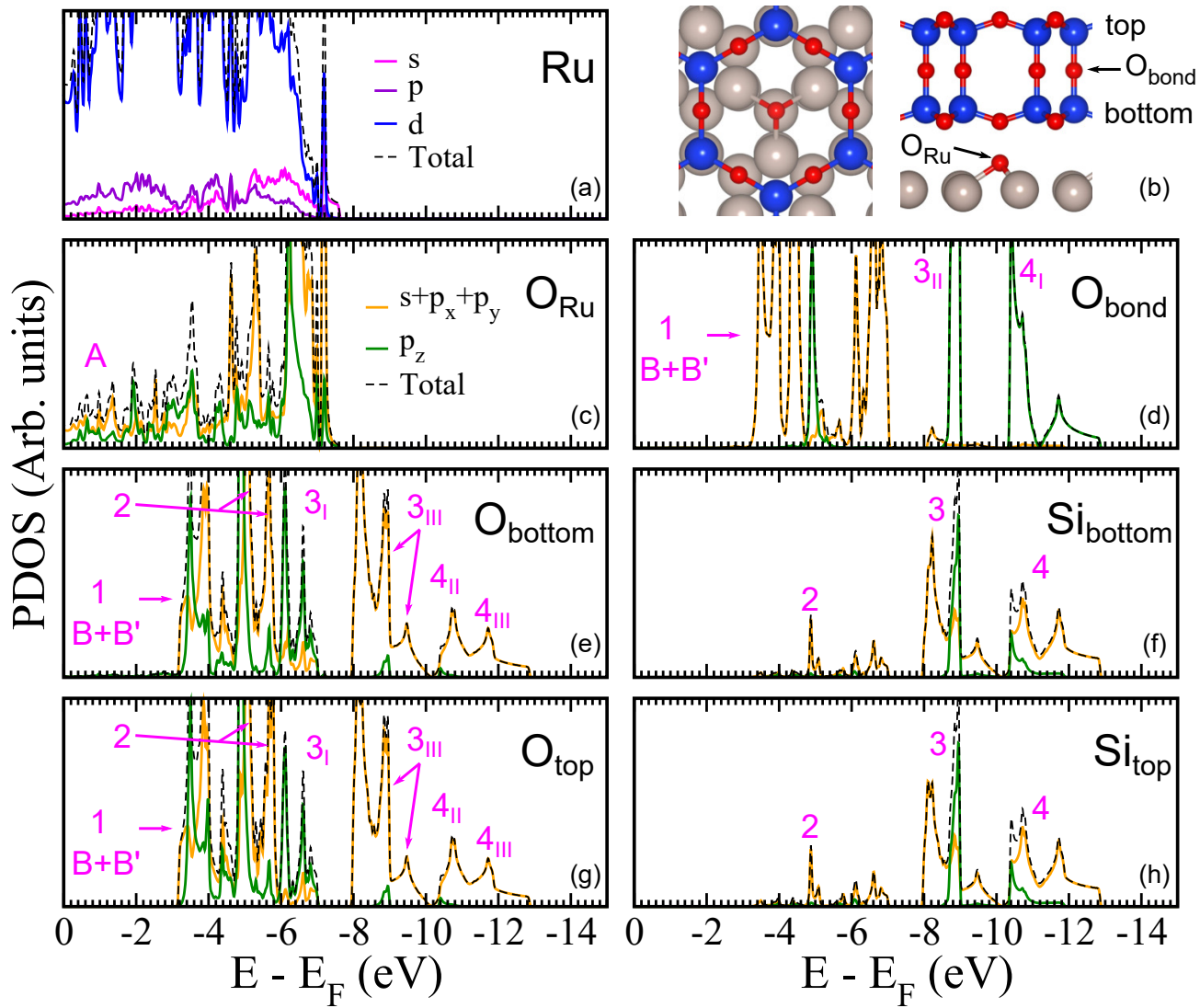

Figure S4: Orbital-projected density of states on Ru, Si and O atoms in 2D BL silicon oxide. Atoms are labeled according to the structural scheme displayed in panel (b). Ru, Si and O atoms are represented in grey, blue and red respectively.

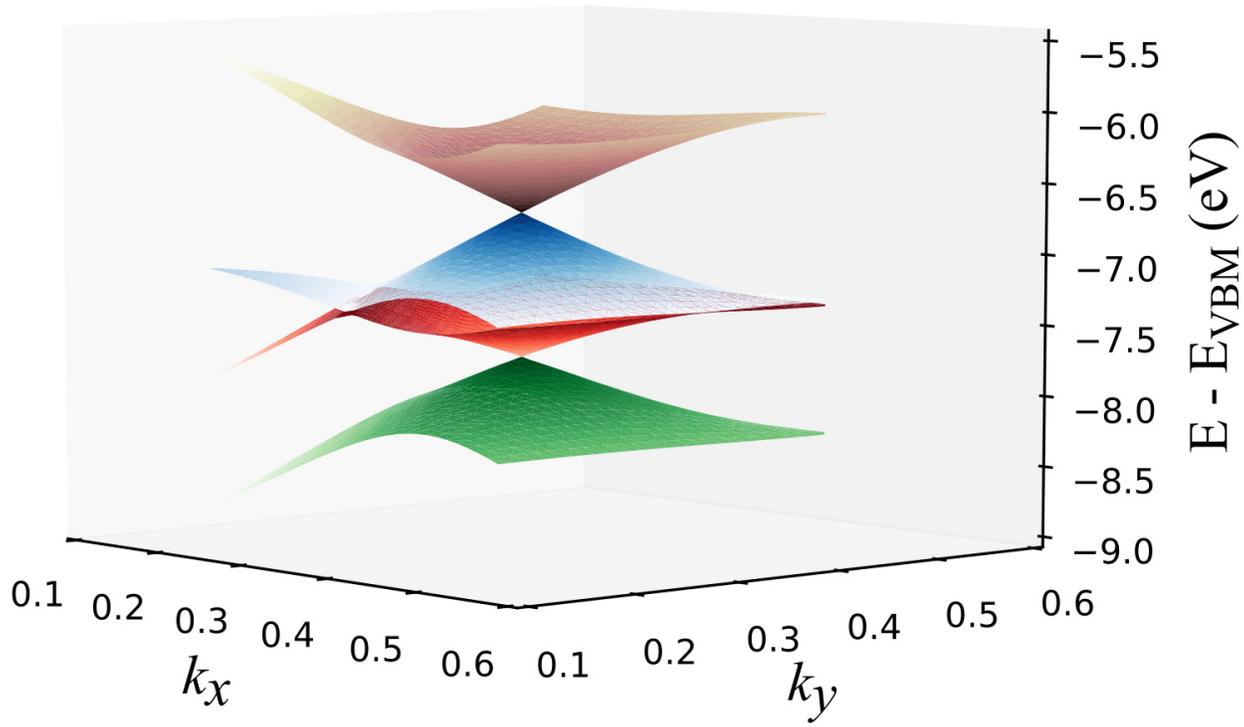

Figure S5: Three-dimensional band structure showing the linear dispersions around the CP1 and CP2 points (as defined in figure 4(c)) as calculated with the LDA functional for the free-standing BL silicon oxide. Brillouin zone points (1/3, 1/3) and (0, 1/2) correspond to the symmetric K and M ones respectively.

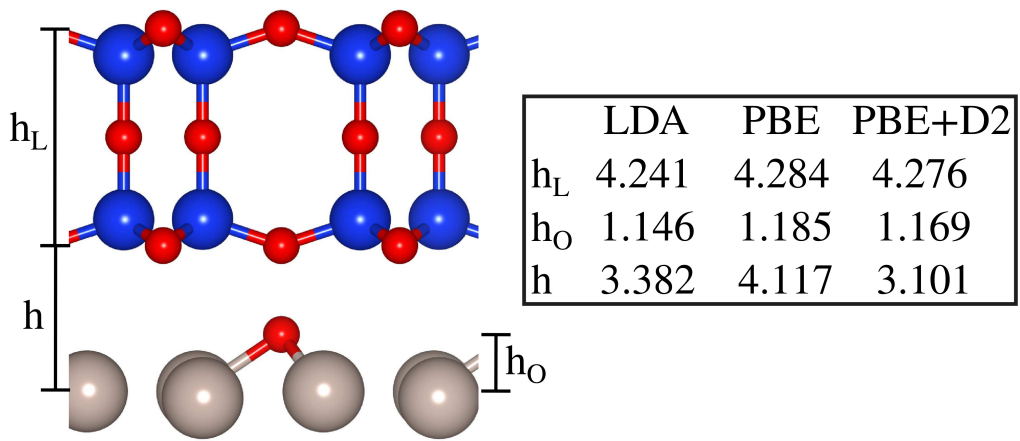

Figure S6: Left: Schematic side view of the 2D BL silicon oxide on Ru(0001). BL silicon oxide width ($h_L$), its height ($h$) with respect to the top Ruthenium layer and the height of the oxygen atom bonded to the Ru surface ($h_O$) are indicated. Right: $h_L$, $h_O$ and $h$ values calculated with the LDA, PBE and PBE+D2 functionals. Distances are given in Å.

For the study of the BL silicon oxide supported on the Ru(0001) surface, we used, in spite of the LDA funcional, the generalized gradient approximation (GGA) with the PBE parametrization[1] on which we included the van der Waals (vdW) correction through the D2 method of Grimme.[2] Figure S6 shows a comparison of the BL silicon oxide width ($h_L$) as well as the height of the oxygen atom bonded to the surface ($h_O$) and the BL height ($h$) calculated as the distance between the average position of the atoms on the top ruthenium layer and that of the oxygen atoms at the BL bottom. PBE and PBE+D2 slightly changes both the isolated BL silicon oxide and oxidized ruthenium surface subsystems with respect to the LDA values. However, the BL–Ru surface distance, characterized by $h$, strongly depends on the functional. Its largest value is obtained with the PBE functional while the smallest one is produced by including the vdW correction on it. LDA gives a value close to the one produced by PBE+D2 indicating that the geometrical structure is well described by the LDA functional.

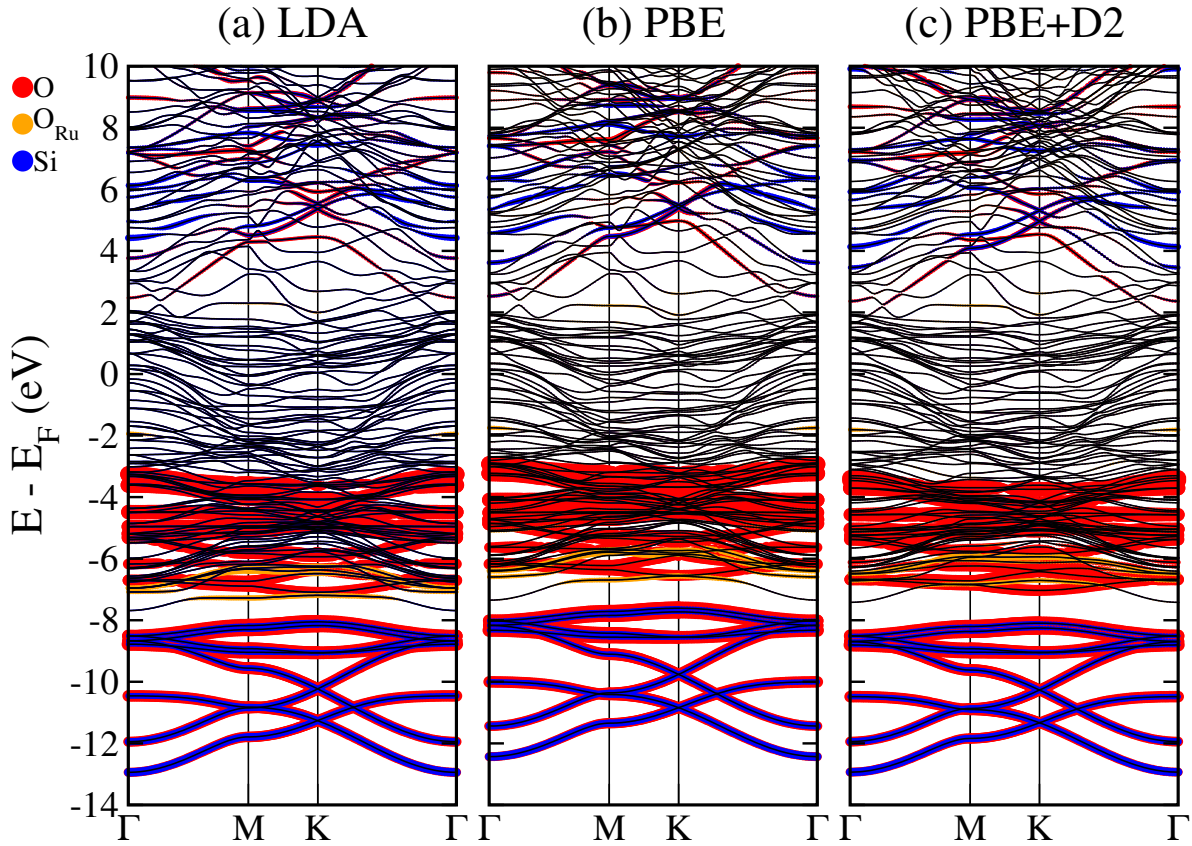

Figure S7: Electronic band dispersion of BL silicon oxide on Ru(0001) calculated with the (a) LDA, (b) PBE and (c) PBE+D2 functionals. Orange, red and blue dots correspond to the projection on the oxygen atom on top of Ru ($O_{Ru}$), oxygen (O) and silicon (Si) atoms on the BL silicon oxide respectively.

Figure S7 shows a comparison of the electronic band dispersion of the BL silicon oxide supported on the Ru(0001) surface obtained with the LDA, PBE and PBE+D2 functionals. Besides a rigid upward shift of the PBE band structure with respect to that of LDA, the electronic band dispersion obtained with these both functionals looks similar. By including the vdW correction on the PBE functional, the BL silicon oxide states (blue and red dots) remain unafected in spite of a rigid downward shift with respect to the Ru and Ru-O states.

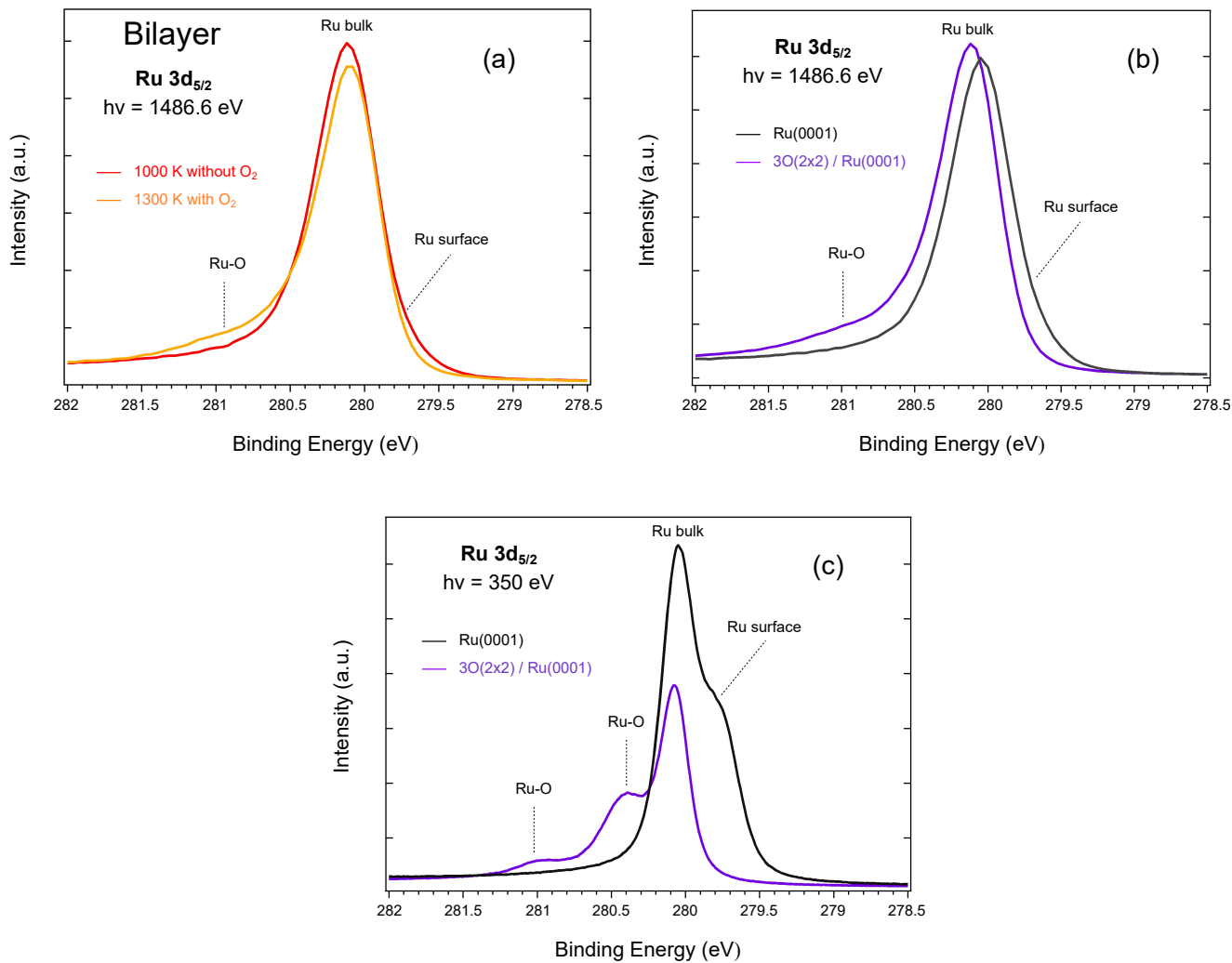

Figure S8: High resolution XPS spectra of the Ru $3d_{5/2}$ core level taken at $h\nu = 1486.6$ eV with a resolution of 300 meV for (a) BL silicon oxide annealed at 1000 K without $O_2$ (red) and 1300 K with $O_2$ (orange); (b) Clean Ru(0001) surface (black) and 3O–(2×2)/Ru(0001) (purple). (c) Same as (b) with $h\nu = 350$ eV. Spectra from this last panel have been measured at the Cassiopee beamline of the synchrotron Soleil with a resolution of 70 meV.

9

# References

1. John P. Perdew, Kieron Burke, and Matthias Ernzerhof. Generalized gradient approximation made simple. *Phys. Rev. Lett.*, 77:3865–3868, Oct 1996.

2. Stefan Grimme. Semiempirical GGA-type density functional constructed with a long-range dispersion correction. *J. Comput. Chem.*, 27(15):1787–1799, 2006.



\end{center}

\end{document}